\documentclass[%
 reprint,
 amsmath,amssymb,
 aps,
]{revtex4-1}

\usepackage{graphicx}
\usepackage{dcolumn}
\usepackage{bm}
\usepackage{bbold}
\usepackage{subfigure}
\usepackage{natbib}

\usepackage{adjustbox}

\usepackage{color}

\DeclareMathOperator{\Tr}{Tr}
\DeclareMathOperator{\Real}{Re}
\DeclareMathOperator{\LL}{\mathcal{L}}

\newcommand*{\bd}{\boldsymbol}

\begin{document}


\title{{Generalized patterns from local and non local reactions}}

%
%

\author{Giulia Cencetti$^{1}$, Federico Battiston$^{2}$, Timoteo Carletti$^{3}$, Duccio Fanelli$^{1}$ \vspace*{.25cm}}
\affiliation{$^1$Dipartimento di Fisica e Astronomia, Universit\`{a} degli Studi di Firenze, INFN and CSDC, Via Sansone 1, 50019 Sesto Fiorentino, Firenze, Italy}
\affiliation{$^2$Department of Network and Data Science, Central European University, Budapest 1051, Hungary}
\affiliation{$^3$naXys, Namur Institute for Complex Systems, University of Namur, Belgium}

\begin{abstract}
{ A class of  systems is considered, where immobile species associated to distinct patches, the nodes of a network, interact both locally and at a long-range, as specified by  an (interaction) adjacency matrix}. Non local {interactions} are treated in a mean-field setting which enables the system to reach a homogeneous consensus state, either constant or time dependent. We provide analytical evidence that such homogeneous solution can turn unstable under externally imposed disturbances, following a symmetry breaking mechanism which anticipates the subsequent outbreak of the patterns. The onset of the instability can be traced back, via a linear stability analysis, to a dispersion relation that is shaped by the spectrum of an unconventional {\it reactive Laplacian}. The proposed mechanism prescinds from the classical Local Activation and Lateral Inhibition scheme, which sits at the core of the Turing {recipe for diffusion driven instabilities}.  Examples of systems displaying a fixed-point or a limit cycle, in their uncoupled versions, are discussed. Taken together, our results pave the way for alternative mechanisms of pattern formation, opening new possibilities for modeling ecological, chemical and physical interacting systems.  

\end{abstract}

\pacs{Valid PACS appear here}

\maketitle


\section{Introduction}

Reaction-diffusion systems are universally invoked as a reference modeling tool, owing to their inherent ability to sustain the spontaneous generation of -- spatially extended -- patterned motifs. The theory of pattern formation was laid down by A. Turing~\cite{Turing52} and later on fertilized in a cross-disciplinary perspective to eventually become a pillar to explain self-organization in nature~\cite{Strogatz01Nature,Pecora_etal97,Pismen06}. In the original Turing setting, two chemicals, here termed species, can relocate in space following Fickean diffusion. In general, a form of local feedback is needed: this activates the short range production of a given species, which should be, at the same time, inhibited at long ranges. In standard reaction-diffusion systems, this combination is accomplished via autocatalytic reaction loops and having the inhibitors diffusing faster than activators. The Local Activation and Lateral Inhibition (LALI) paradigm provides hence the reference frame for the onset of self-organized patterns of the Turing type.

Both in the continuous and discrete (lattice or network based ~\cite{OthmerScriven74,NakaoMikhailov10,AsllaniChallengerPavoneSacconiFanelli14}) versions, diffusion is a key ingredient for the Turing paradigm to apply. Assume in fact that a stable 
fixed-point (or, alternatively a time dependent equilibrium, e.g. a limit cycle) exists for the scrutinized reaction scheme in its a-spatial local version. Then, the spatial counterpart of the inspected model admits a homogeneous equilibrium. This follows trivially by recalling that by definition the Laplacian, the operator that implements the diffusive transport, returns zero when applied to a uniform density background. Tiny erratic disturbances, that happen to perturb the uniform solution, may therefore activate the diffusion term, and consequently feed on the reaction part, in this way promoting a self-sustained instability which sits at the core of the Turing mechanism. 

Natural systems display however an innate drive towards self-organization, which certainly transcends the realms of validity of Turing's ideas. In many cases of interest, the quantities to be monitored are anchored to the nodes of a virtual graph. The information to be processed on site is carried across the edges of the graph. Nevertheless, the spreading of information springing from nearby nodes is not necessarily bound to obey linear diffusion, as in the spirit of the original Turing formulation. At variance, it can in principle comply with a large plethora of distinct dynamical modalities. In ecology, the number of nodes reflects  biodiversity: inter-species interactions are usually epitomized by a quadratic response function, which accounts for  competitive predator-prey or symbiotic dependences~\cite{May72,ThebaultFontaine10}. In genetic networks, inputs between neighboring genes are shaped by a sigmoidal profile, a non-linear step function often invoked to mimic threshold activation processes~\cite{BecskeiSerrano00}. 
At a radically different scale, the dynamics of opinions in sizable social systems are customarily traced back to pairwise exchanges, which can echo physical encounters or be filtered by device-assisted long-ranged interactions~\cite{castellano2009statistical}. Similarly, global pandemic events mirror complex migration patterns, which cannot be explained as resorting to the simplistic archetype of diffusion~\cite{pastor2001epidemic}. 

A first insight into the study of pattern formation for a system made of immobile species was offered in ~\cite{BullaraDecker15}. In this latter paper the colored motifs on the skin of a zebrafish were reproduced by postulating an experimentally justified long ranged regulatory mechanism, in absence of cell motility. The differential growth mechanism identified by Bullara and De Decker~\cite{BullaraDecker15} yields a LALI paradigm alternative to that stemming from conventional reaction diffusion schemes.   

Motivated by these observations, we here take one leap forward by proposing a generalized theory of pattern formation which is not bound to satisfy the LALI conditions. {Interactions are assumed to be 
mediated} by local (on site) and non local (distant, possibly long-ranged) pairings among species. More specifically, we will assume that non local interactions, as specified by a off-diagonal elements of a binary or weighted adjacency matrix, can be scaled by the node degree, following a mean-field working ansatz.  The elemental units of the system being inspected (species) are permanently linked to the nodes they are bound to, and therefore classified as immobile.  {Local interactions will be also referred to as reactions, to establish an ideal bridge with standard reaction-diffusion systems.}

\section{The scheme of interaction}

Consider a reactive system made of two, mutually interacting species and label with $x$ and $y$ their respective concentrations. Assume: 
\begin{equation}
	\left\{\begin {split}
	\dot x =& f(x,y) \\
	\dot y =& g(x,y).\\
	\end{split}\right.
	\label{unc_sys}
\end{equation}
where $f(\cdot,\cdot)$ and $g(\cdot,\cdot)$ are generic non-linear functions.  Suppose that the system admits an equilibrium solution, $(x^*,y^*)$, which can be either a fixed-point or a 
limit cycle. Linearizing equations (\ref{unc_sys}) around  ${\bd w^*}= (x^*,y^*)$ returns a (linear) first order system for the evolution of the imposed perturbation $\delta {\bd w} = (\delta x, \delta y)^T$, where $T$ stands for the transpose operation. In formulae, one gets $\delta \dot{\bd w} =   \bd{J} (\bd w^*) \delta{\bd w}$, where $\bd{J}(\bd w^*)=\left(
\begin{array}{ c c}
\partial_xf^* & \partial_yf^* \\ 
\partial_xg^* & \partial_yg^* 
\end{array}
\right)$ 
is the Jacobian matrix of the system evaluated at equilibrium. When the equilibrium is a fixed-point, the stability of ${\bd w^*}$ is set by the spectrum of $\bd{J}$: if the largest real part of the eigenvalues of $\bd{J}$ is negative the imposed perturbation fades away exponentially and the fixed-point is deemed stable. This amounts to requiring $\Tr(\bd J)<0$ and $\det(\bd J)>0$, a pre-requisite condition for our analysis to apply.  When the equilibrium solution is a limit cycle, the Jacobian exhibits a periodic dependence on time and the stability can be assessed by evaluating the Floquet exponents~\cite{Grimshaw17,ChallengerBurioniFanelli15,LucasFanelliCarletti18}. In the following, we develop our reasoning by assuming a constant Jacobian. Our analysis and conclusions, however, readily extend to a limit cycle setting, provided one replaces $\bd{J}$ with the time-independent Floquet matrix.

Starting from a local formulation of the reactive model, we move forward and account for non local interactions. We begin by replicating the examined system on $N$ different patches, defining  lumps of embedding space or specific criteria to effectively group the variables involved. The collection of $N$ isolated spots where the 
local {interaction} scheme is replicated are the nodes of a network, identified by the index $i=1,...,N$. The edges of the network provide a descriptive representation of the non-local {interaction} scheme: $A_{ij}=1$, if nodes $i$ and $j$  share a link, {i.e. if they are bound to mutually interact} (the formalism extends readily to account for weighted interactions). For the sake of simplicity we limit the analysis to undirected networks, $A_{ij}=A_{ji}$.
We note that, by definition, we have $A_{ii}=1$. The connectivity of node $i$ is hence $k_i^A=\sum_j A_{ij}$.
Rewrite the non-linear function $f$ as $f(x,y)=f_0(x,y)+f_1(x,y)$, where $f_0$ speaks for the interaction terms that are genuinely local, i.e. those not modulated by distant couplings. Conversely, $f_1(x,y)$ will be replaced by a  reaction function, which encodes for the remote interaction. In the spirit of a nearest neighbors mean-field approximation, we shall substitute $f_1(x,y)$ with its averaged counterpart in the heterogeneous space of the interaction network.  Similarly, $g$ will be split as $g(x,y)=g_0(x,y)+g_1(x,y)$. Then we set: 

\begin{equation}
\label{system1}
	\left\{\begin {split}
	\dot x_i =& f_0(x_i,y_i) + \frac{1}{k_i^A}\sum_jA_{ij}\tilde{f}_1(x_i,x_j,y_i,y_j)\\
	\dot y_i =& g_0(x_i,y_i) + \frac{1}{k_i^A}\sum_jA_{ij}\tilde{g}_1(x_i,x_j,y_i,y_j).
	\end{split}\right.
\end{equation}
where $\tilde{f}_1$  (resp. $\tilde{g}_1$) reduces to $f_1$  (resp. $g_1$) when the system is in a homogeneous state. Mathematically, we require
$\tilde{f}_1(x_i,x_j,y_i,y_j) \delta_{ij}  = f_1(x_i,y_i)$ and $\tilde{g}_1(x_i,x_j,y_i,y_j) \delta_{ij} = g_1(x_i,y_i)$. As an illustrative example, consider a local quadratic interaction of the type 
$f_1(x_i)=x_i^2$. This latter can be turned into $\tilde{f}_1(x_i,x_j)=x_i x_j$, a choice that matches the above constraints. This would yield $(x_i/k_i^A) \sum_j A_{ij} x_j = x_i \langle x \rangle_i$ in the first of equations (\ref{system1}), where the average $\langle \cdot \rangle_i$ runs over the nodes adjacent to $i$.  In practical terms, as anticipated above, we replace the local reaction term $f_1(x_i,y_i)$ to include interactions at distance, where the average runs over the  $k_i^A$ terms defining the neighborhood of $i$. The remote interaction is assumed to depend only on the species that populate the selected pair of nodes.

By adding and  subtracting  $f_1(x_i,y_i)$ (resp. $g_1(x_i,y_i)$ ) in the first (resp. second) of equations (\ref{system1}) and making use of the definition of $f(\cdot,\cdot)$ and $g(\cdot,\cdot)$ yields:
 
\begin{equation}
	\left\{\begin {split}
	\dot x_i = & f(x_i,y_i) + \sum_j\LL_{ij}^{\rm R}\tilde{f}_1(x_i,x_j,y_i,y_j)\\
	\dot y_i = & g(x_i,y_i)+ \sum_j\LL_{ij}^{\rm R}\tilde{g}_1(x_i,x_j,y_i,y_j)\\
	\end{split}\right.
\label{syst_eq}
\end{equation}
where  $\LL_{ij}^{\rm R} = \frac{A_{ij}}{k_i^A} - \delta_{ij}$. Observe that the above equations are formally identical to system (\ref{unc_sys}) when $\bd A \equiv \mathbb{1}_N$, where $\mathbb{1}_N$ stands for the $N \times N$ identity matrix. In this case, in fact, the $N$ nodes are formally decoupled and  system (\ref{syst_eq}) reduces to (\ref{unc_sys}), for each node of the ensemble. System 
(\ref{syst_eq}) admits an obvious homogeneous solution. This latter is found by assigning $(x_i,y_i)=(x^*,y^*)$ $\forall i$. Can one make the 
homogeneous equilibrium unstable to tiny non homogeneous perturbations by acting on the network of interactions, as encoded in the operator $\bd{{\LL}}^{\rm R}$? {Answering this question amounts to develop a general theory for pattern formation for interacting system, in absence of species diffusion.}  
To expand along this line one needs to characterize the operator $\bd{{\LL}}^{\rm R}$. As shown in the Appendix, it is easy to prove that $\bd{{\LL}}^{\rm R}$ displays a non-positive spectrum and that has an eigenvector $(1,\dots,1)^T$ associated to the $0$ eigenvalue. This justifies referring to $\bd{{\LL}}^{\rm R}$ as to a {\it reactive Laplacian}. In the literature it is also known as the  consensus Laplacian~\cite{Ghosh_etal14,Lambiotte_etal11,OlfatiSaber07,Krause09}. It is important to notice that this is  different from the diffusive Laplacian  ($L^{diff}_{ij} = A_{ij} - k_j^A\delta_{ij}$) and its normalized version ($L^{RW}_{ij} = \frac{A_{ij}}{k_j^A} - \delta_{ij}$) which describes random walk. {It is not possible to reduce the reactive Laplacian to the Laplacian of diffusion, unless the complex network reduces to a simple regular lattice.}


\section{Linear stability analysis}

The stability of the homogeneous equilibrium $(x_i,y_i)=(x^*,y^*)$, $\forall i$ can be analytically probed  {\cite{pecora1998master}}.  Introduce
a small inhomogeneous perturbation, $x_i=x^*+u_i$, $y_i=y^*+v_i$, and expand the governing equations up to the leading (linear) order to yield: 
\begin{equation}
\left(
\begin{array}{ c }
\dot{\bd u}\\ 
\dot{\bd v}
\end{array}
\right) 
= \bd{J} \otimes \mathbb{1}
\left(
\begin{array}{ c }
 \bd u\\ 
 \bd v
\end{array}
\right)
+ \bd{J}_R \otimes \bd{{\LL}}^{\rm R}
\left(
\begin{array}{ c }
 \bd u\\ 
 \bd v
\end{array}
\right)
\label{pert_syst}
\end{equation}
where $\bd u=\left(u_1,u_2, ..., u_N \right)$, $\bd v=\left(v_1,v_2, ..., v_N \right)$, $\bd{J}_R = \left(
\begin{array}{ c c}
\partial_2\tilde{f}_1^* & \partial_4\tilde{f}_1^* \\ 
\partial_2\tilde{g}_1^* & \partial_4\tilde{g}_1^* 
\end{array}
\right)$. The symbols $\partial_2$ and $\partial_4$ indicate that the derivatives are computed with respect to the second and fourth arguments of the function to which they are applied. The symbol $\otimes$ refers to the Kronecker product. 
More details on the derivation of eq. (\ref{pert_syst}) are given in the Appendix. To solve the above linear system, we introduce the eigenvectors ($\phi_j^{(\alpha)}$) and eigenvalues ($\Lambda^{(\alpha)}$) of the reactive Laplacian $\bd{{\LL}}^{\rm R}$, i.e. $\sum_j\LL_{ij}^{\rm R}\phi_j^{(\alpha)}=\Lambda^{(\alpha)}\phi_i^{(\alpha)}$ with $\alpha=1,..., N$. Expanding the perturbations on the basis of  $\phi_j^{(\alpha)}$, $\alpha=1,..., N$,  we obtain (see Appendix) the  $2 \times 2$ modified Jacobian 
$\bd{J}_{\alpha} =\bd{J} + \Lambda^{(\alpha)}\bd{J}_R$, which determines the stability of the system.  For each choice of  $\Lambda^{(\alpha)}$,  one needs to compute $\lambda(\alpha)$, the largest real part of the eigenvalues of $\bd{J}_{\alpha}$.   The discrete collection of  $\lambda(\alpha)$ vs. $-\Lambda^{(\alpha)}$ defines the so-called dispersion relation. The homogeneous equilibrium is stable if  $\lambda(\alpha) < 0$, $\forall \alpha$. Conversely, the perturbation grows if at least one $\lambda(\alpha)$ turns out to be positive. In this latter case patterns can emerge, thus breaking the symmetry of the unperturbed initial condition. The condition of instability can be expressed in a compact form (provided $\bd{J}^{-1}$ exists), as we prove in the Appendix. Specifically, we find that the instability sets in if one of the following conditions holds: (i) if $\Tr(\bd{J}_R)<0$  and provided  $\Lambda^{(\alpha)} < -\frac{\Tr(\bd{J})}{\Tr{\bd{J}_R}}$, for at least one choice of $\alpha$; (ii) if $\det(\bd{J}_R)<0$ and provided  $\Lambda^{(\alpha)} <  \Lambda_{-}$, for at least one choice of $\alpha$;
(iii)\ if $\det(\bd{J}_R)>0$, $\Tr(\bd{J}^{-1}\bd{J}_R)>0$ and provided  $\Lambda_-<\Lambda^{(\alpha)} <  \Lambda_+$, for at least one choice of $\alpha$. Here, $\Lambda_{\pm} = \frac{1}{2}\biggl[ -\Tr(\bd{J}^{-1}\bd{J}_R)\pm \sqrt{[\Tr(\bd{J}^{-1}\bd{J}_R)]^2 - 4\det(\bd{J}^{-1}\bd{J}_R)}\biggr]$. The above conclusions 
hold for a two species model that displays a constant homogeneous equilibrium. {As such, it } generalizes the renowned Turing instability conditions for reaction-diffusion models, to  systems  {made of immobile species} subject to local and  {non local (remote)} interactions. The analysis extends to the setting where the homogeneous state is a synchronous collection of oscillators performing in unison. As we shall also remark in the following, the novel route to pattern formation overcomes the LALI constraint.

{Before turning to discussing the applications, we remark that generalized transport schemes have been considered in the past which enable one to relax the LALI request. This includes for instance accounting for cross-diffusion 
\cite{vanag2009cross,madzvamuse2015cross}. It should be emphasized however that within the current scheme, species are immobile, hence permanently bound to their reference nodes. It is in fact the combination of local and non-local interaction rules that materializes in the discrete non-local operator, different from the Laplacian of diffusion, which rules the dynamics of the system in its linear approximation.}

\begin{figure}[htb]
\centering
 \begin{adjustbox}{center}
\includegraphics[width=1.05\columnwidth]{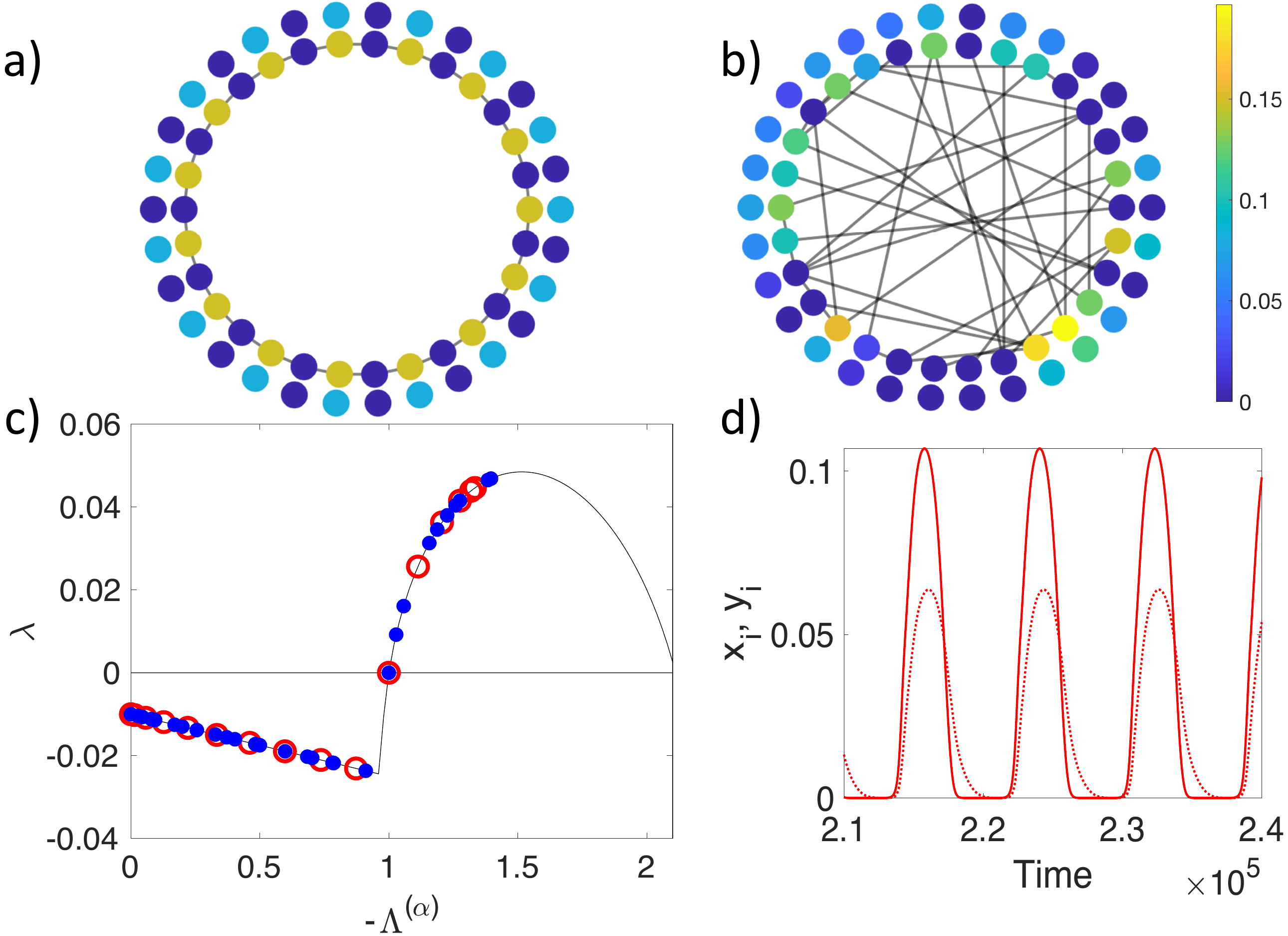}
 \end{adjustbox}
 \caption{Panels (a) and (b): node colors code for the asymptotic  stationary concentration of the species (predators, outer ring, and prey, inner ring), for the Volterra model for two distinct interaction networks. 
 Panel (c): the associated dispersion relations $\lambda(\alpha)$ vs. $-\Lambda^{(\alpha)}$. Red symbols refer to the one-dimensional lattice with periodic boundary conditions, and blue symbols follow from a ER network. 
Panel (d):  for specific choices of the parameters, different from those adopted in panels a) and b) , the random inhomogeneous perturbations of the homogeneous equilibrium give raise to predator (dashed line)-prey (continuous line) oscillations, testifying on the rich gallery of generated patterns.   
}
\label{fig:fig1}
\end{figure}

\section{Applications and discussions}

As a first example, we consider the reaction scheme known as the Volterra model~\cite{mckane2005predator}, $f(x,y)=  c_1 xy - d x$ and  $g(x,y)=ry - sy^2 - c_2xy$.
The variable $x$ identifies the concentration of predators, while $y$ stands for the prey. The parameters are positive definite. The Volterra equations admit a fixed-point, $x^* = \frac{c_1r-sd}{c_1c_2}$, $y^*=\frac{d}{c_1}$, which is meaningful and stable provided $c_1r-sd>0$. We now turn to considering $N$ replicas of the model, by making the variables to depend on the node index $i$, associated to different ecological niches. Following the above lines of reasoning, we assume
that species can sense the remote interaction with other communities populating the neighboring nodes. For instance, the competition of prey for food and resources can be easily 
extended so as to account for a larger habitat which embraces adjacent patches. At the same time, predators can benefit from a coordinated action to hunt in team. Assume that the linear terms in $f(x,y)$ and $g(x,y)$ stay local (death term for predators and birth term for prey), i.e. $f_0(x,y)=-dx$, $g_0(x,y)=ry$. We generalize the terms  $f_1(x,y)=c_1xy$ and $g_1(x,y)=-sy^2-c_2xy$ to the case of remote interactions as:
\begin{equation}
	\left\{\begin {split}
	\dot x_i =& -d x_i + \beta \frac{c_1}{k_i} y_i\sum_j A_{ij}x_j + (1-\beta)\frac{c_1}{k_i}x_i\sum_j A_{ij}y_j\\
	\dot y_i =& ry_i - \frac{s}{k_i}y_i\sum_j A_{ij}y_j - \frac{c_2}{k_i}y_i\sum_j A_{ij}x_j\\
	\end{split}\right.
	\label{Volterra}
\end{equation}
where  $\beta \in (0,1)$  modulates the relative strength of the non local terms that drive the evolution of the predators. By simple manipulations, we cast the previous equations so as to bring into evidence the dependence on the reactive Laplacian $\bd{{\LL}}^{\rm R}$. In formulae, one gets

\begin{eqnarray}
\nonumber
\dot x_i &=& c_1 x_iy_i - d x_i + \alpha c_1y_i\sum_j \LL^{\rm R}_{ij}x_j + (1-\alpha)c_1x_i\sum_j\LL^{\rm R}_{ij}y_j\\ \nonumber
\dot y_i &=& ry_i - sy_i^2 - c_2x_iy_i - sy_i\sum_j\LL^{\rm R}_{ij}y_j - c_2y_i\sum_j\LL^{\rm R}_{ij}x_j. 
\end{eqnarray}

The system can then be studied by resorting to the strategy developed above. The interaction parameters can be set so as to make the dispersion relation positive,  as displayed in Fig. \ref{fig:fig1}(c). Red symbols refer to the generalized Volterra model organized on a close lattice made of $N=30$ nodes (panel (a) of Fig.~\ref{fig:fig1}), while the blue symbols are obtained when the sites are connected by an Erd\H{o}s-R\'enyi (ER) network with $\langle k \rangle = 3.3$  (panel (b)). The colors of the nodes stand for the asymptotic stationary state of the system across the networks, showing the formation of distinct patterns for predators (outer ring) and prey (inner ring), in the two cases. While the lattice yields a regular, equally spaced, pattern, a disordered distribution of populated patches is instead obtained for the ER network, thus reflecting the randomness of the connections. We remark that in both cases, the extinction in a patch of one of the species is typically associated to the disappearance of the other one in the same node, resulting in a global habitat with populated patches alternated with uninhabited ones. These latter 
play the role of natural barriers separating colonized niches. The emerging patterns of coexistence differ from those observed from standard reaction-diffusion systems, where predators and prey tend to cluster in distinct sites.
The mutually exclusive distribution of species, as obtained in classical reaction diffusion scheme, bears the imprint of the LALI assumption. This latter does not apply for  the patterns reported in Fig 1.  As a matter of fact, the associated Jacobian matrix displays  non positive diagonal entries.
As a side result, we notice that in some cases, Fig. \ref{fig:fig1}(d),  the perturbation may trigger the emergence of regular oscillations, typical of a predator-prey cycle. This behavior cannot be reproduced for a deterministic Volterra model in absence of remote couplings~\cite{mckane2005predator}.
\begin{figure}[htb]
\centering
 \begin{adjustbox}{center}
\includegraphics[width=1.05\columnwidth]{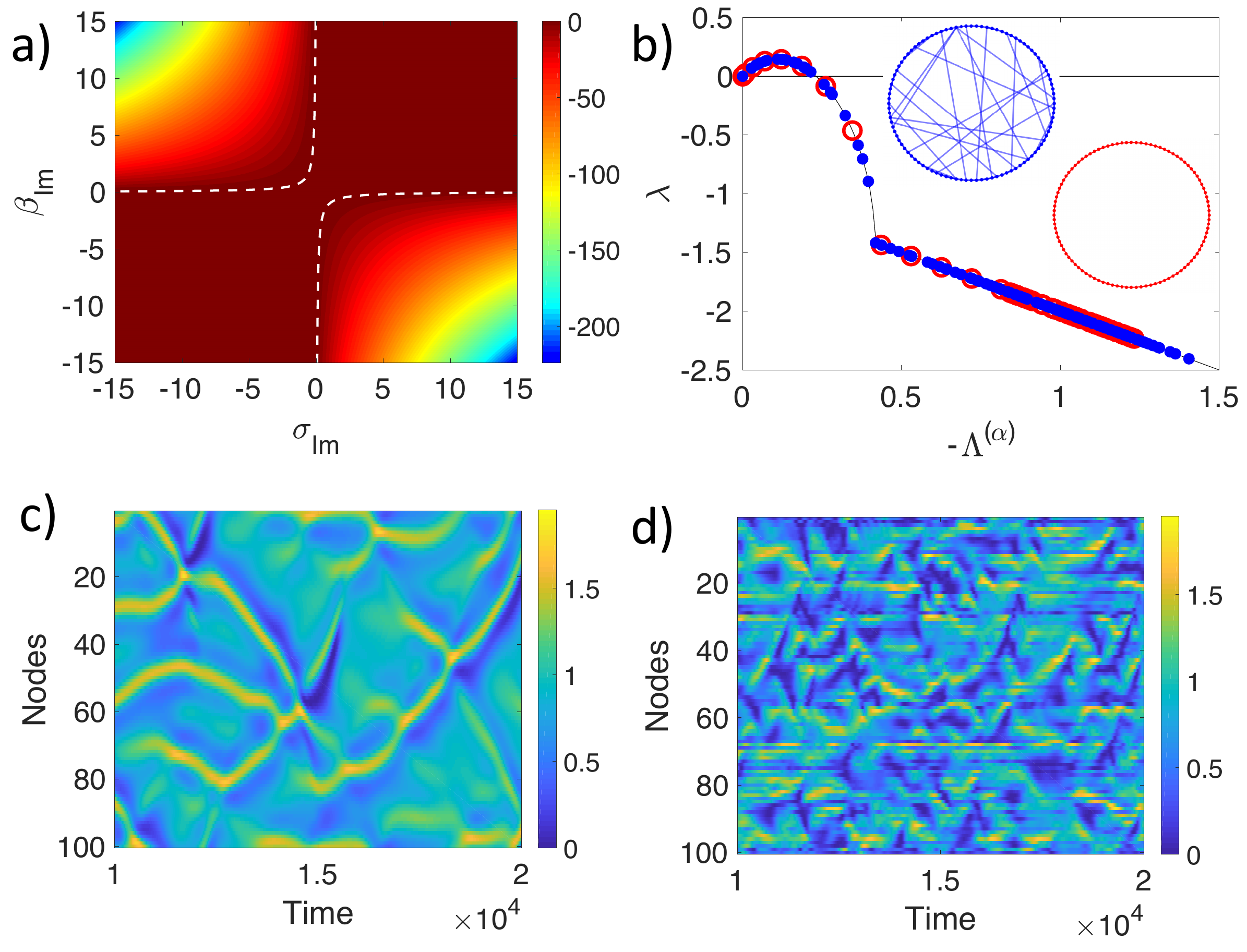}
 \end{adjustbox}
 \caption{In (a) the instability domain is depicted:  $c=\sigma_{Re}+\sigma_{Im}\beta_{Im}/\beta_{Re}$ is plotted at varying $\sigma_{Im}$ and $\beta_{Im}$, while $\sigma_{Re}=\beta_{Re}=1$. The instability condition for the limit cycle $w^*(t)$ coincides with the request $c<0$. Panel (b) shows the dispersion relation for two different networks, a lattice with $k=6$ (red empty circles) and a Watts-Strogatz network with the same mean degree (blue full circles). Panels (c) and (d) show the patterns of $|w_i|$, in the space of the nodes and against time, corresponding to the previously mentioned network structures and ensuing dispersion relations.}
\label{fig:fig2}
\end{figure}

As a second example, we  consider a set of  {interacting} Stuart-Landau oscillators \cite{BordyugovPikovskyRosenblum10,Teramae04,Selivanov_etal12,PanaggioAbrams15,Zakharova_etal14}.  The Stuart-Landau equation takes the form $\dot w = \sigma w - \beta w |w|^2$, where $w=w_{Re} + i w_{Im}$ is a complex number, as well as the parameters $\sigma, \beta$.
The above equation yields a stable limit cycle which can be cast in the explicit form $w^*(t) = \sqrt{\frac{\sigma_{Re}}{\beta_{Re}}}\exp(i \omega t)$, where 
$\omega = \sigma_{Im} - \beta_{Im} \sigma_{Re}/\beta_{Re}$. To determine the stability of the time dependent periodic solution $w^*$, one can introduce a perturbation of $w^*$ in the form $w^*\left(1+u \right) \exp(  v)$, where $u$ and $v$ are small quantities. Expanding at the linear order of approximation yields a Jacobian matrix 
$\bd{J} =
\left(
\begin{array}{ c c}
-2\sigma_{Re} & 0 \\ 
-2\beta_{Im} \sigma_{Re}/\beta_{Re} & 0 
\end{array}
\right).$
Interestingly, even if $w^*$ is periodic in time, $\bd{J}$ is constant, owing to the specificity of the Stuart-Landau equation and  to the form of the imposed perturbation. This observation allows us to proceed in the analysis without resorting to the Floquet machinery. The request of a stable limit cycle $w^*$ translates into $\sigma_{Re}>0$, which consequently implies $\beta_{Re}>0$ (since $\sigma_{Re} / \beta_{Re}>0$ for the solution to exist). Stuart-Landau oscillators are often coupled diffusively, either on a regular lattice or on a complex network, so as to result in the celebrated Ginzburg-Landau equation. For suitable choices of the parameters, the oscillators evolve in unison, giving rise to a fully synchronized homogenous state. This latter can however break apart, upon injection of a tiny disturbance, if the parameters are assigned in the complementary reference domain. This modulational instability, known in the literature as the Benjamin-Feir instability, follows from the subtle interplay between reaction and diffusion terms~\cite{DiPattiFanelliCarletti16,BenjaminFeir67}. Inspired by the analysis carried out above, here we consider a  collection of non diffusive Stuart-Landau oscillators, interacting via their linear reaction term. This choice materializes in a variant of the traditional Ginzburg-Landau equation, ensuing a novel instability of the Benjamin-Feir type. Consider an ensemble made of $i=1,...,N$ oscillators at the nodes of a network, characterized by the complex amplitude $w_i$, with $i=1,...,N$. The family of Stuart-Landau oscillators evolves therefore as:
$\dot w_i  = \frac{\sigma}{k_i} \sum_jA_{ij} w_j - \beta w_i |w_i|^2$ or, equivalently, 
$\dot w_i = \sigma w_i - \beta w_i |w_i|^2 + \sigma \sum_j \LL_{ij}^{\rm R} w_j$.
 A linear stability analysis which moves from the ansatz $w_i=w^*\left(1+u_i \right) \exp(  v_i)$ (with $u_i,v_i << 1$) and implements the above calculation strategy leads to the  
$\bd{J}_{\alpha} = \bd{J} + \Lambda^{(\alpha)}\bd{J}_R$, where 
 $\bd{J}_R =
\left(
\begin{array}{ c c}
\sigma_{Re} & -\sigma_{Im} \\ 
\sigma_{Im} & \sigma_{Re}
\end{array}
\right).$
As $\Tr(\bd{J}_R) = 2 \sigma_{Re}>0$, the homogeneous solution turns unstable only if $\det(\bd J_{\alpha})<0$. It is  easy to prove that $\det(\bd J_{\alpha})<0$ given  $\Lambda^{(\alpha)} > 2\frac{\sigma_{Re}^2+\sigma_{Re}\sigma_{Im}\beta_{Im}/\beta_{Re}}{\sigma_{Re}^2+\sigma_{Im}^2}$, for at least one choice of $\alpha$. Since the eigenvalues of the reactive Laplacian are by definition negative and $\sigma_{Re}>0$, the instability request materializes in the compact necessary condition $\sigma_{Re}+\sigma_{Im}\frac{\beta_{Im}}{\beta_{Re}}<0$, which constitutes the revisitation of the Benjamin-Feir instability to the present setting. Fig. \ref{fig:fig2}(a) displays the region of instability in the reference plane ($\sigma_{Im}, \beta_{Im}$) (when setting $\sigma_{Re}$ and $\beta_{Re}$ to unit). The dashed white line marks the transition to the region of instability. Emerging patterns, and their associated dispersion relations, $\lambda(\alpha)$ vs. $-\Lambda^{(\alpha)}$, are depicted for the Stuart-Landau oscillators distributed on, respectively, a lattice with non-local couplings and a Watts-Strogatz graph. In the Appendix, we show that long-range couplings promotes the stability of the system. 

Summing up,  we have here introduced a novel mechanism for the emergence of coherent patterns  {for systems made of interacting, although immobile, species}.  Remote interactions, treated in a mean-field approximation, lead to a reactive Laplacian, whose spectrum ultimately sets the conditions for the instability. Taken all together, our theory paves the way for investigating novel mechanisms of pattern formation, by providing a flexible alternative to traditional schemes that implements the LALI constraint.

	\addcontentsline{toc}{chapter}{Bibliography}
	\bibliographystyle{apsrev4-1} 
	\bibliography{bib_Turing}

\onecolumngrid


\appendix


\subsection{Characterizing the spectrum of the reactive Laplacian.}

Recalling that $A_{ii}=1$, let us introduce matrix $\bd{{\mathcal{A}}}=\bd A-\mathbb{1}$, the adjacency matrix of the same network, after removal of the self-loops. Hence, $k_i^\mathcal{A}=\sum_j \mathcal{A}_{ij}=k_i^A-1$ for all $i$. Define then $L_{ij}^A=A_{ij}-k_i^A\delta_{ij}$ and $L_{ij}^\mathcal{A}=\mathcal{A}_{ij}-k_i^\mathcal{A}\delta_{ij}$. We can straightforwardly prove that $L_{ij}^A=L_{ij}^\mathcal{A}$. Indeed,
$L_{ii}^A=A_{ii}-k_i^A=1-k_i^A=-k_i^\mathcal{A}=L_{ii}^\mathcal{A}$ and $\forall i\neq j$ $L_{ij}^A=A_{ij}=\mathcal{A}_{ij}=L_{ij}^\mathcal{A}$.
$L_{ij}^\mathcal{A}$ is by definition a Laplacian matrix, hence characterised by a non-positive spectrum. Based on the above the same conclusion applies to $L_{ij}^A$. In particular, $(1,\dots,1)^T$ is the eigenvector of $\bd L^A$ associated to the $0$ eigenvalue. We now turn to considering the reactive Laplacian $\bd{{\LL}}^{\rm R}$ and set to prove that it also displays a non-positive spectrum, thus providing an a posteriori justification to the name assigned to the operator. Denote by $\bd D$ the diagonal matrix of the connectivities of $\bd{{\LL}}^{\rm R}$, $D_{ii}=k_i^A$. Then, introduce $\bd L^{sym}=\bd D^{-1/2}\bd L^A\bd D^{-1/2}$. By using the definition of $\bd L^A$ we get $L^{sym}_{ij}=A_{ij}/\sqrt{k_i^A k_j^A} -\delta_{ij}$ from which it immediately follows that $\bd L^{sym}$ is symmetric. We can then show that $\bd L^{sym}$ is non-positive definite. Take any $\bd x\in\mathbb{R}^N\setminus \{0\}$, $N$ standing for the dimension of the matrices, i.e. the size of the network. Then $(\bd x,\bd L^{sym} \bd x)=(\bd x,\bd D^{-1/2}\bd L^A\bd D^{-1/2} x)=(\bd D^{-1/2}\bd x,\bd L^A\bd D^{-1/2} \bd x)$
where we used the property of the scalar product $(\bd x,\bd M\bd x)=(\bd M^T\bd x,\bd x)$ and the fact the $\bd D^{-1/2}$ is diagonal. By setting $\bd y=\bd D^{-1/2} \bd x$, we  get
$(\bd x,\bd L^{sym} \bd x)=(\bd y,\bd L^A\bd y)\leq 0\quad \forall \bd x$
where the last inequality follows from the fact that $\bd L^A$ is non-positive definite. Hence $\bd L^{sym}$ shares the same property. We are finally in a  position to show that $\bd{{\LL}}^{\rm R}$ has the characteristic spectrum of a
Laplacian matrix. Observe that $\bd{{\LL}}^{\rm R}=\bd D^{-1}\bd L^A =\bd D^{-1/2}\left(\bd D^{-1/2}\bd L^A\bd D^{-1/2}\right)\bd D^{1/2}=\bd D^{-1/2}\bd L^{sym}\bd D^{1/2}$.
Hence, $\bd{{\LL}}^{\rm R}$ is similar to $\bd L^{sym}$ and, thus, the two operators  display the same non-positive spectrum. Finally, by the very first definition, it follows that $(1,\dots,1)^T$ is the eigenvector associated to the $0$ eigenvalue of $\bd{{\LL}}^{\rm R}$, which ends the proof.

\subsection{The dispersion relation and the conditions for instability}

In the following we provide a detailed procedure to characterize the linear stability of the homogeneous equilibria $(x^*,y^*)$ for systems in the general form of eqs.~(2) in the main text.\\ 
First of all, we introduce a small inhomogeneous perturbation, $x_i=x^*+u_i$, $y_i=y^*+v_i$, and expand the governing equations up to the linear order to eventually get: 
\begin{equation}
	\left\{\begin {split}
	\dot u_i = & \partial_x f^*u_i + \partial_y f^*v_i + \sum_j\LL_{ij}^{\rm R}[\partial_2f_R^*u_j + \partial_4f_R^*v_j]\\
	\dot v_i = & \partial_x g^*u_i + \partial_y g^*v_i + \sum_j\LL_{ij}^{\rm R}[\partial_2g_R^*u_j + \partial_4g_R^*v_j]
	\end{split}\right.
\end{equation}
where the symbols $\partial_2$ and $\partial_4$ indicate that the derivatives are computed with respect to the second and fourth arguments of the function to which they are applied. In matrix form, 

\begin{equation}
\left(
\begin{array}{ c }
\dot{u_i}\\ 
\dot{v_i}
\end{array}
\right) 
= \bd{J} 
\left(
\begin{array}{ c }
  u_i\\ 
  v_i
\end{array}
\right)
+ \sum_j \LL_{ij}^{\rm R} \bd{J}_R 
\left(
\begin{array}{ c }
 u_j\\ 
 v_j
\end{array}
\right)
\label{pert_syst0}
\end{equation}

where 
\begin{equation}
\bd{J}_R =
\left(
\begin{array}{ c c}
\partial_2f_R^* & \partial_4f_R^* \\ 
\partial_2g_R^* & \partial_4g_R^* 
\end{array}
\right).
\label{syst_eq_lin1}
\end{equation}

Eq.~\eqref{pert_syst0} can be written in a more compact form by making use of the Kronecker product, $\otimes$:

\begin{equation}
\left(
\begin{array}{ c }
\dot{\bd u}\\ 
\dot{\bd v}
\end{array}
\right) 
= \bd{J} \otimes \mathbb{1}
\left(
\begin{array}{ c }
 \bd u\\ 
 \bd v
\end{array}
\right)
+ \bd{J}_R \otimes \bd{{\LL}}^{\rm R}
\left(
\begin{array}{ c }
 \bd u\\ 
 \bd v
\end{array}
\right)
\label{pert_syst1}
\end{equation}

Introduce the eigenvectors and eigenvalues of the Laplacian $\bd{{\LL}}^{\rm R}$, as $\sum_j\LL_{ij}^{\rm R} \phi_j^{(\alpha)} = \Lambda^{(\alpha)}\phi_i^{(\alpha)}$. Expand the perturbations on the basis of the reactive Laplacian to eventually yield:
\begin{equation}
	u_i=\sum_{\alpha}c_{\alpha}\phi_i^{(\alpha)}, \hspace{8mm}
	v_i=\sum_{\alpha}d_{\alpha}\phi_i^{(\alpha)}.
\end{equation}
We can therefore write equation~\eqref{pert_syst1} as 
\begin{equation}
\sum_{\alpha} \left(
\begin{array}{ c }
\dot{c_{\alpha}}\\ 
\dot{d_{\alpha}}
\end{array}
\right) \bd \phi^{(\alpha)}
= \bd{J} \otimes \mathbb{1}
\sum_{\alpha} \left(
\begin{array}{ c }
 c_{\alpha}\\ 
 d_{\alpha}
\end{array}
\right) \bd \phi^{(\alpha)}
+ \bd{J}_R \otimes \bd{{\LL}}^{\rm R}
\sum_{\alpha}  \left(
\begin{array}{ c }
 c_{\alpha}\\ 
 d_{\alpha}
\end{array}
\right) \bd \phi^{(\alpha)},
\end{equation}
and by diagonalizing the reactive Laplacian:
\begin{equation}
\sum_{\alpha} \left(
\begin{array}{ c }
\dot{c_{\alpha}}\\ 
\dot{d_{\alpha}}
\end{array}
\right) \bd \phi^{(\alpha)}
= [\bd{J} + \Lambda^{(\alpha)} \bd{J}_R] \otimes \mathbb{1}
\sum_{\alpha} \left(
\begin{array}{ c }
 c_{\alpha}\\ 
 d_{\alpha}
\end{array}
\right) \bd \phi^{(\alpha)}.
\end{equation}
The above linear system is  ruled by a $2\times N$ block-diagonal Jacobian matrix. Each block is the $2\times 2$ reduced Jacobian $\bd{J}_{\alpha} \equiv \bd{J} + \Lambda^{(\alpha)}\bd{J}_R$.
By making use of  the fact that the eigenvectors are linearly independent one eventually gets system 

\begin{equation}
\left(
\begin{array}{ c }
\dot{c_{\alpha}}\\ 
\dot{d_{\alpha}}
\end{array}
\right) 
= [ \bd{J} + \Lambda^{(\alpha)}\bd{J}_R ]
\left(
\begin{array}{ c }
 c_{\alpha}\\ 
 d_{\alpha}
\end{array}
\right).
\label{reduced}
\end{equation}

\begin{figure}[ht]
\includegraphics[width=.4\textwidth]{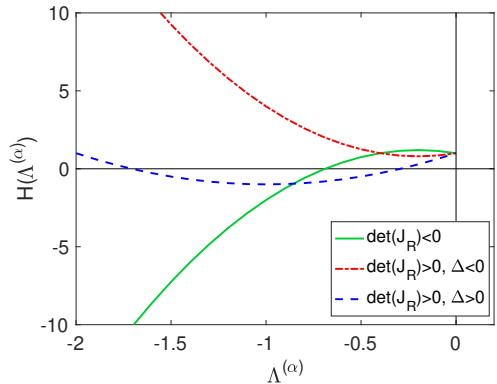}
\caption{The function $H(\Lambda^{(\alpha)})$ is displayed against $\Lambda^{(\alpha)}$ to highlight the three different scenarios that can be eventually met.}
\label{fig:SI}
\end{figure}

Starting from this setting we can elaborate on the conditions that drive the system unstable.  
The eigenvalues of the reduced Jacobian $ \bd{J}_{\alpha}$ are given by:
\begin{equation}
\lambda_{\pm}(\alpha) = \frac{1}{2}\biggl[ \Tr(\bd{J}_{\alpha} )\pm \sqrt{\Tr(\bd{J}_{\alpha} )^2 - 4\det(\bd{J}_{\alpha})}\biggr].
\end{equation}
It is straightforward to verify that, in absence of couplings, i.e. for $\Lambda^{(\alpha)}=0$, the real part of  $\lambda_{\pm}$ is always negative, as it should be because of the stability of the uncoupled equilibrium. We now seek to find the condition for which there exists at least one eigenvalue of the reactive Laplacian, that corresponds to a positive entry of the dispersion relation, i.e. $\Real(\lambda_{\pm})>0$, so signaling the instability. This condition is clearly met  if $\Tr(\bd{J}_{\alpha} ) > 0$ or if $\det(\bd{J}_{\alpha} ) < 0$.\\
The first condition can be written as $\Tr(\bd{J}) +  \Lambda^{(\alpha)}\Tr(\bd{J}_R) > 0$. By recalling that the Laplacian eigenvalues are negative and that $\Tr(\bd J)<0$, one finds that the latter condition is matched only if $\Tr(\bd{J}_R)<0$, for  $\Lambda^{(\alpha)} < -\frac{\Tr(\bd{J})}{\Tr{\bd{J}_R}}$.\\
The second condition can be reformulated as $\det(\bd{J})\det(\mathbb{1}+\Lambda^{(\alpha)} \bd{J}^{-1}\bd{J}_R) < 0$, if $\bd J$ is invertible, which is equivalent to requiring $\det(\mathbb{1}+\Lambda^{(\alpha)} \bd{J}^{-1}\bd{J}_R) = {\Lambda^{(\alpha)}}^2\det(\bd{J}^{-1}\bd{J}_R) + \Lambda^{(\alpha)}\Tr(\bd{J}^{-1}\bd{J}_R) +1 \equiv H(\Lambda^{(\alpha)}) < 0$. By observing that $\det(\bd{J}^{-1}\bd{J}_R) = \frac{\det(\bd{J}_R)}{\det(\bd{J})}$ and that $\det(\bd{J}) > 0$, we can isolate two distinct scenarios:
 if $\det(\bd{J}_R) < 0$, the curve $H(\Lambda^{(\alpha)})$ draws a parabola like the one depicted with a  green line in Fig~\ref{fig:SI}. The sought condition is hence matched only if there exists an eigenvalue of the reactive Laplacian smaller than the value where the (continuous) parabola intercepts the $x$-axis.
If instead $\det(\bd{J}_R) > 0$, the curve $H(\Lambda^{(\alpha)})$ is negative defined over a finite interval (blue curve in fig~\ref{fig:SI}), provided  $\Tr(\bd{J}^{-1}\bd{J}_R)>0$ and, at the same time, $\Delta \equiv [\Tr(\bd{J}^{-1}\bd{J}_R)]^2 - 4 \det(\bd{J}^{-1}\bd{J}_R)>0$. It is however easy to prove that the latter inequality is always true. Indeed, one can always write $\Delta=[\Tr(\bd \Gamma)]^2 - 4 \det(\bd \Gamma)$ and, by denoting with $\mu_1, \mu_2$ the eigenvalues of the generic matrix $\bd \Gamma$, we can write $\Delta = (\mu_1+\mu_2)^2 - 4\mu_1\mu_2 = (\mu_1-\mu_2)^2 $, which is always positive.\\

In summary, the homogeneous equilibrium $(x^*,y^*)$ can be unstable in the following cases:

\begin{itemize}
\item if $\Tr(\bd{J}_R)<0$  and provided  $\Lambda^{(\alpha)} < -\frac{\Tr(\bd{J})}{\Tr{\bd{J}_R}}$, for at least one choice of $\alpha$. 
\item if $\det(\bd{J_{\alpha}})<0$ which, if $\bd J$ is invertible, is verified in the following alternative cases:
	\begin{itemize}
	\item if $\det(\bd{J}_R)<0$ and provided  $\Lambda^{(\alpha)} <  \Lambda_{-}$, for at least one choice of $\alpha$

	\item  if $\det(\bd{J}_R)>0$, $\Tr(\bd{J}^{-1}\bd{J}_R)>0$ and provided  $\Lambda_-<\Lambda^{(\alpha)} <  \Lambda_+$, for at least one choice of $\alpha$,
	\end{itemize}
\end{itemize}
where $ \Lambda_{\pm} $ identifies the points where the curve $H(\Lambda^{(\alpha)})$ is equal to zero, namely:
\begin{equation}
 \Lambda_{\pm} = \frac{1}{2}\biggl[ -\Tr(\bd{J}^{-1}\bd{J}_R) \pm \sqrt{[\Tr(\bd{J}^{-1}\bd{J}_R)]^2 - 4\det(\bd{J}^{-1}\bd{J}_R)}\biggr].
\end{equation}

\subsection{Stationary state and Laplacian eigenvectors}

The Laplacian spectrum plays a pivotal role in setting the conditions for the emergence of the patterns, as shown in the previous section. Hereafter, we will show that the spectral properties of the Laplacian convey important information on the characteristics of the patterns that are established in the asymptotic regime. In order to solve the perturbed equations we have indeed expanded the perturbations on the Laplacian eigenvectors basis, thus obtaining 
\begin{equation}
\left(
\begin{array}{ c }
u_i(t)\\ 
v_i(t)
\end{array}
\right) 
= \sum_{\alpha}
\left(
\begin{array}{ c }
 c_{\alpha}(0)\\ 
 d_{\alpha}(0)
\end{array}
\right)
e^{\lambda_{\pm}(\alpha)t} \phi_i^{(\alpha)}.
\end{equation}

\begin{figure}[ht]
\includegraphics[height=6cm]{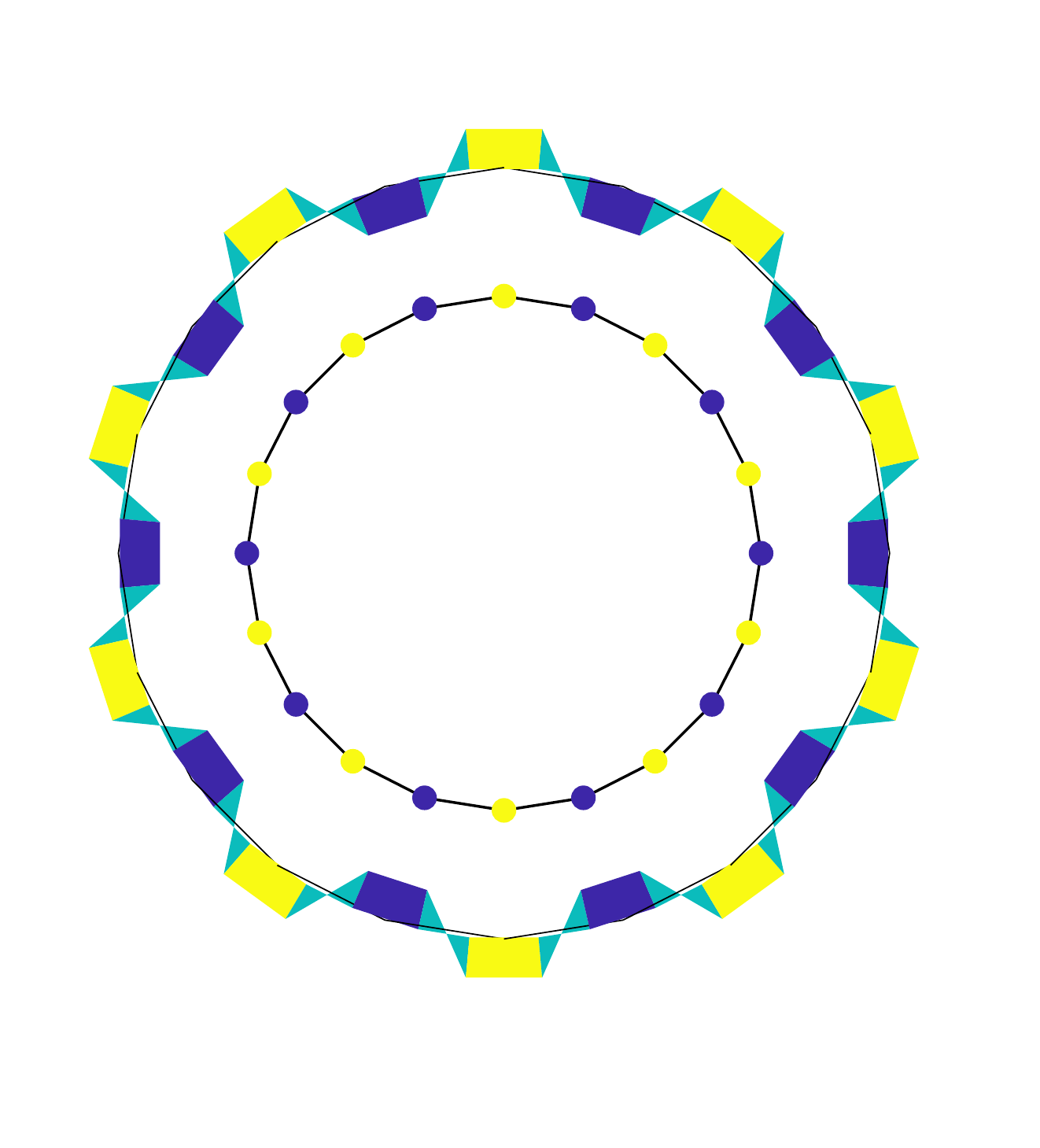}
\includegraphics[height=6cm]{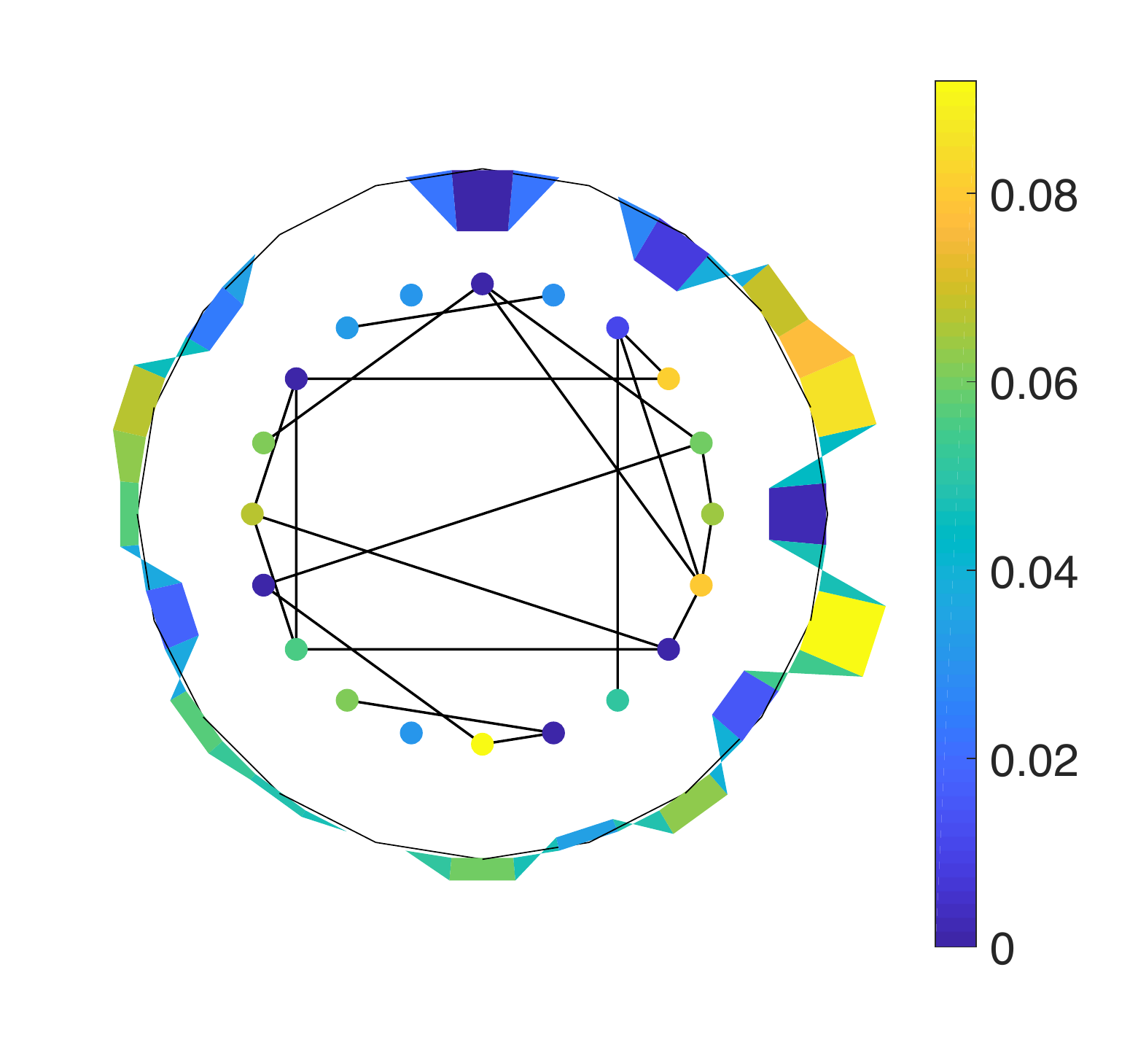}
\caption{The emerging pattern for a Volterra model on a regular lattice and an ER network of 20 nodes are shown. The nodes are colored as follows the stationary density displayed by the species of predators, $x$. The outer ring reports the entries of the most unstable Laplacian eigenvector. The correlation between asymptotic density distribution and ensuing patterns is evident.}
\label{SI_eig}
\end{figure}

The perturbations are therefore projected along independent directions which are biased to reflect the peculiar structure of the network architecture, analytically associating to each direction a growth/damping rate given by the respective entry of the dispersion relation.
The stationary state will be mainly shaped, at the leading order of approximation, by the most unstable mode, the one that displays the larger dispersion relation entry, as it can be
appreciated from figure~\ref{SI_eig}.


\subsection{Network topology affecting pattern formation: the case of the coupled Stuart-Landau equations.}

As an addition to the analysis carried out in the main body of the paper, we here get back to considering the system of coupled Stuart-Landau oscillators and elaborate on the role of long-range interactions for the onset of the instability.  To this end we start from a lattice with $k-1$ nearest neighbors and perturb it with the inclusion of remote links assigned with a given probability. In particular, the generic element $A_{ij}$ is one with probability $|i-j|^{-\gamma}$, where $|\cdot|$ measures the shortest distance on the circle between node $i$ and node $j$. The rewiring is implemented is such a way that the average connectivity is preserved. The exponent $\gamma$ quantifies the range of the interaction, $\gamma_c=1$  setting the threshold between short and long ranged systems, in one dimension.

\begin{figure}

	\subfigure[]{\includegraphics[width=6cm]{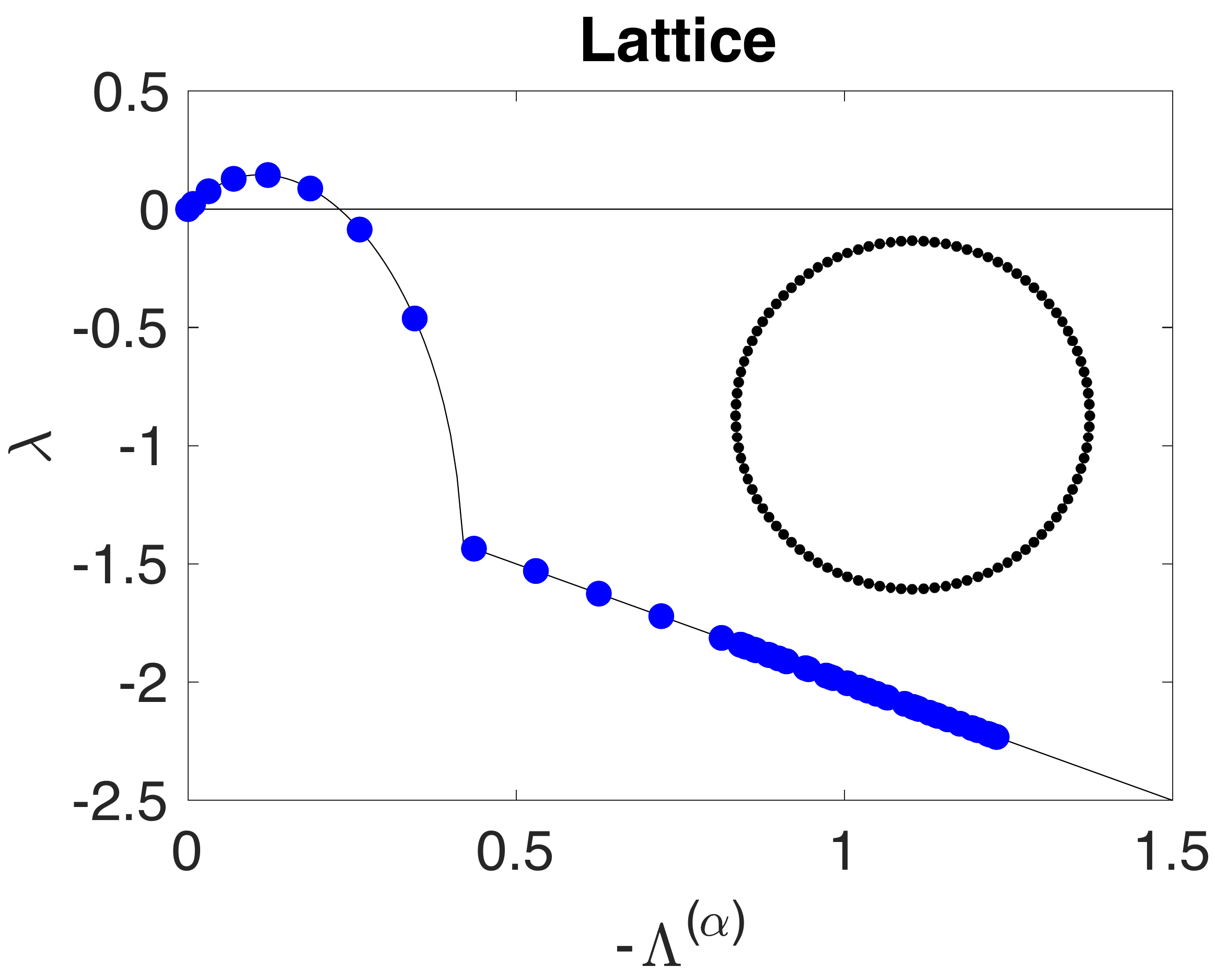}
	\label{1_lattice_dispRel}}
	\hspace{-5mm}
	\subfigure[]{\includegraphics[width=6cm]{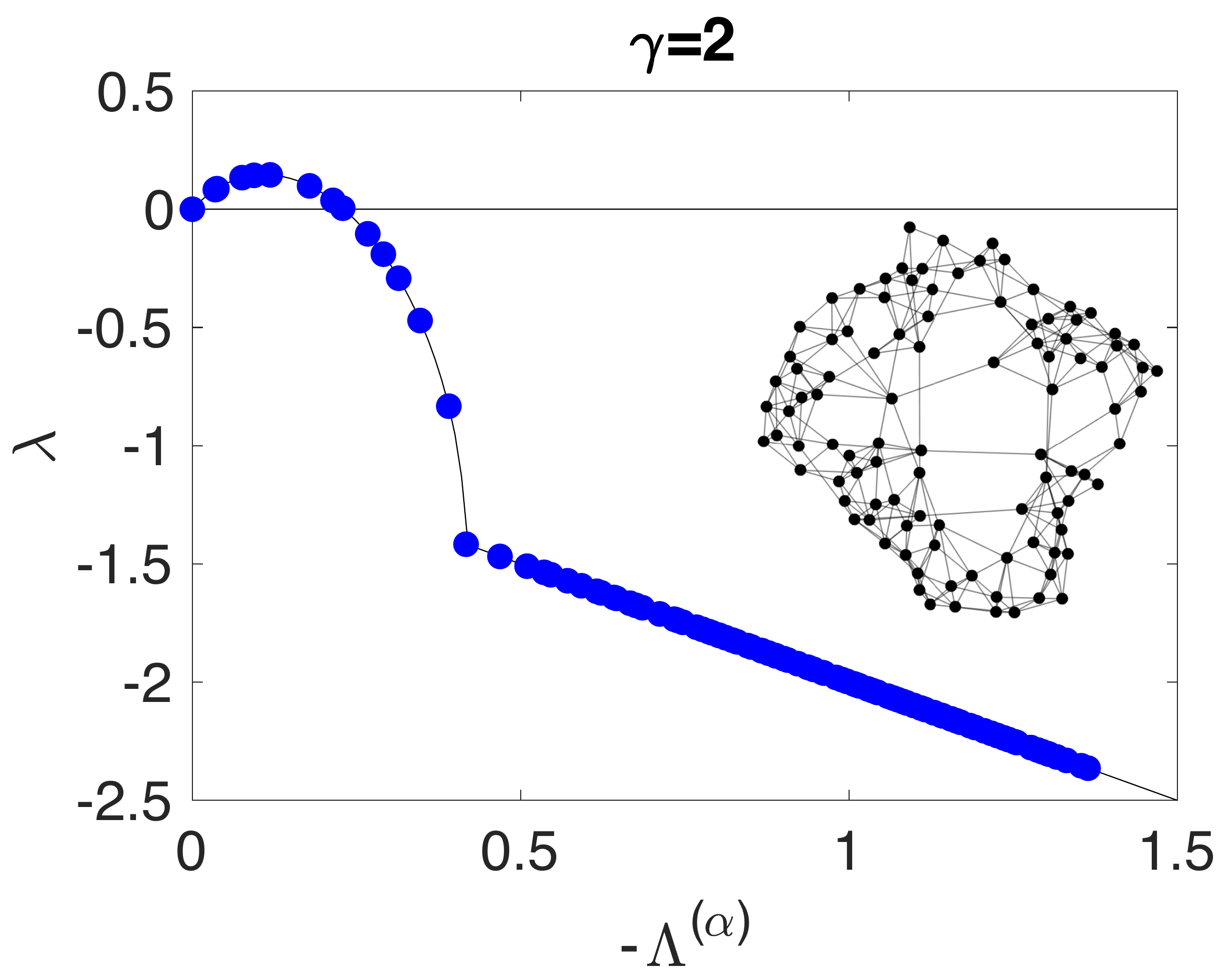}
	\label{1_sr_dispRel}}
	\hspace{-5mm}
	\subfigure[]{\includegraphics[width=6cm]{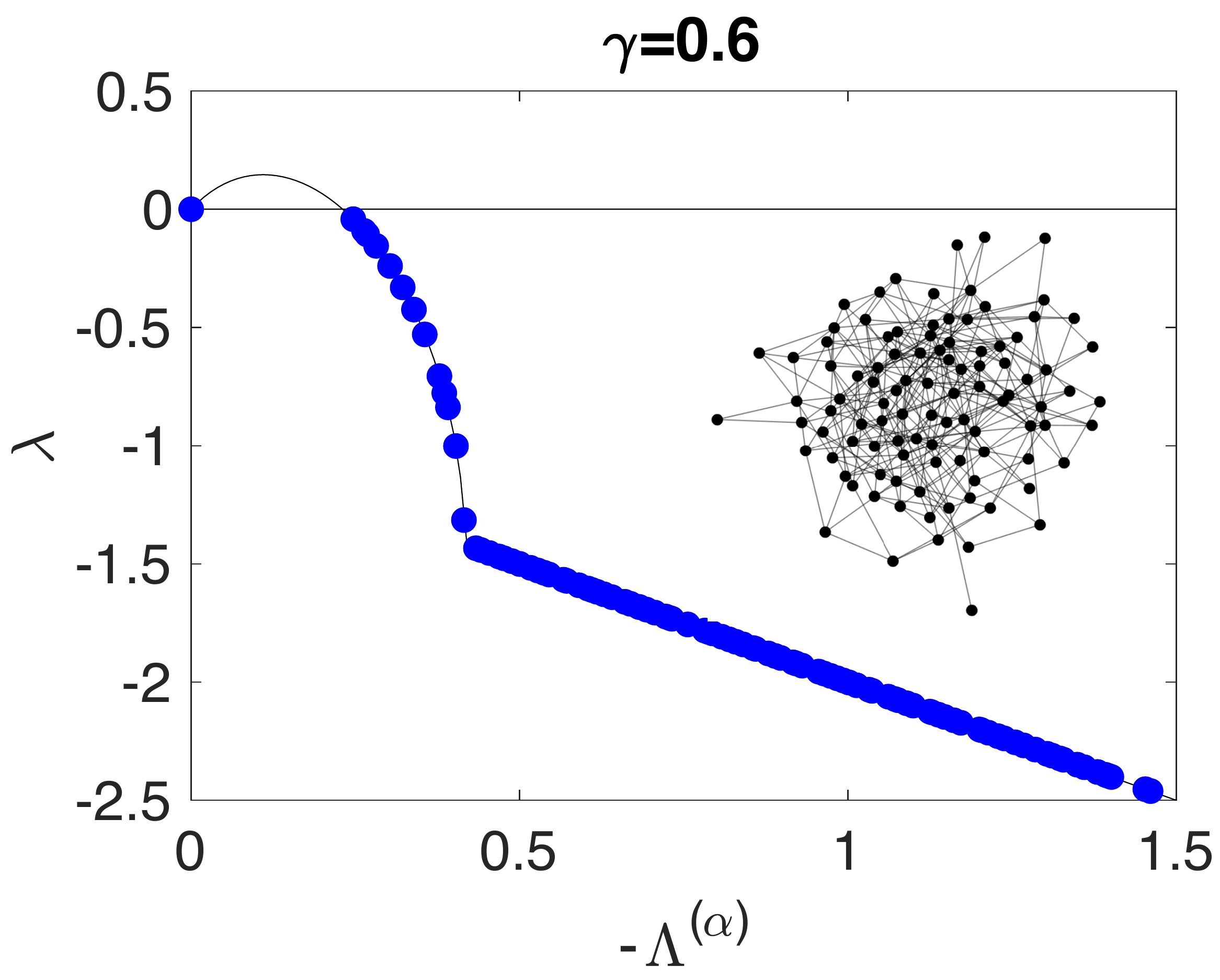}
	\label{1_lr_dispRel}}

	\subfigure[]{\includegraphics[height=4.8cm]{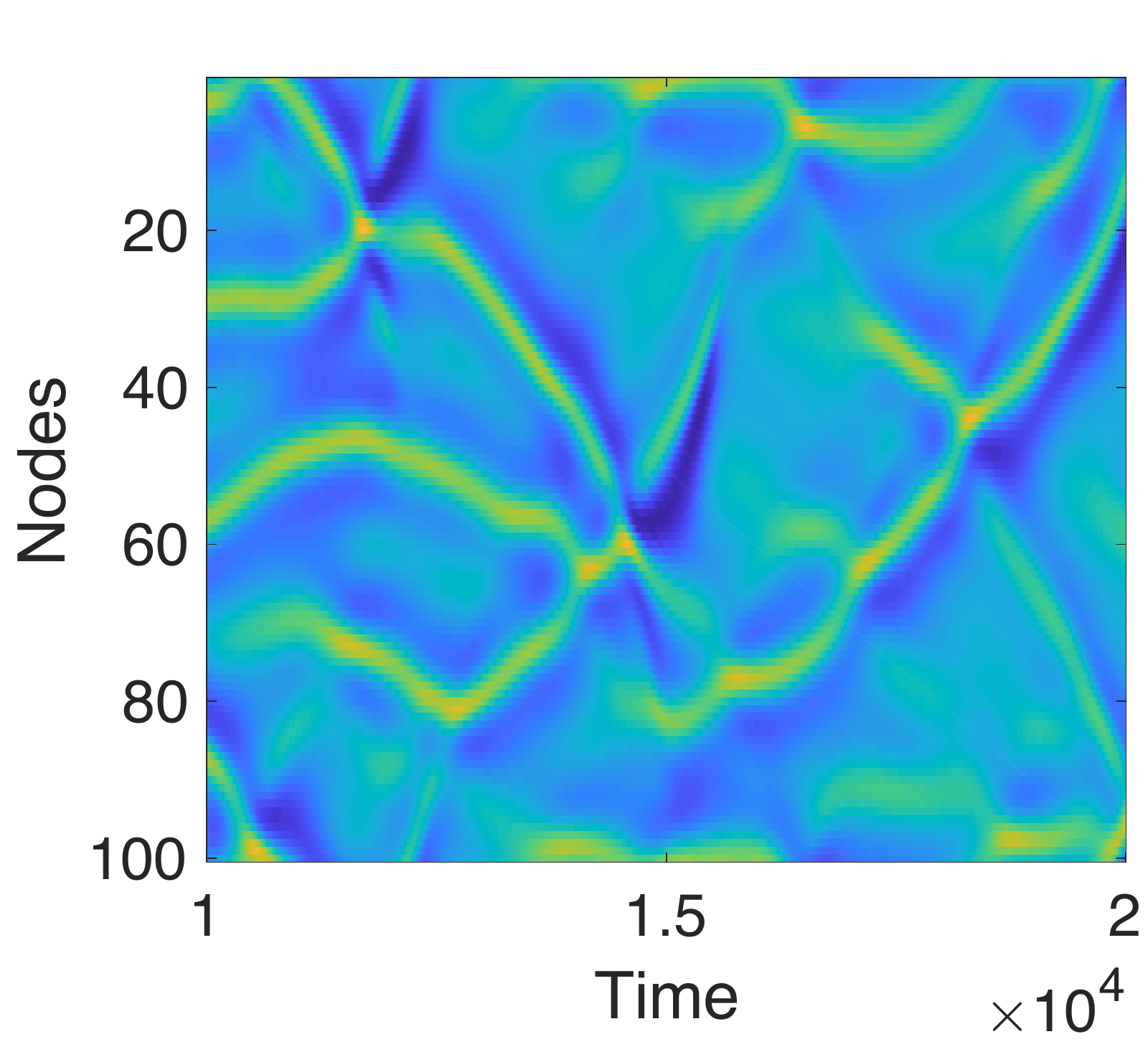}
	\label{1_lattice_mod}}
	\hspace{2mm}
	\subfigure[]{\includegraphics[height=4.8cm]{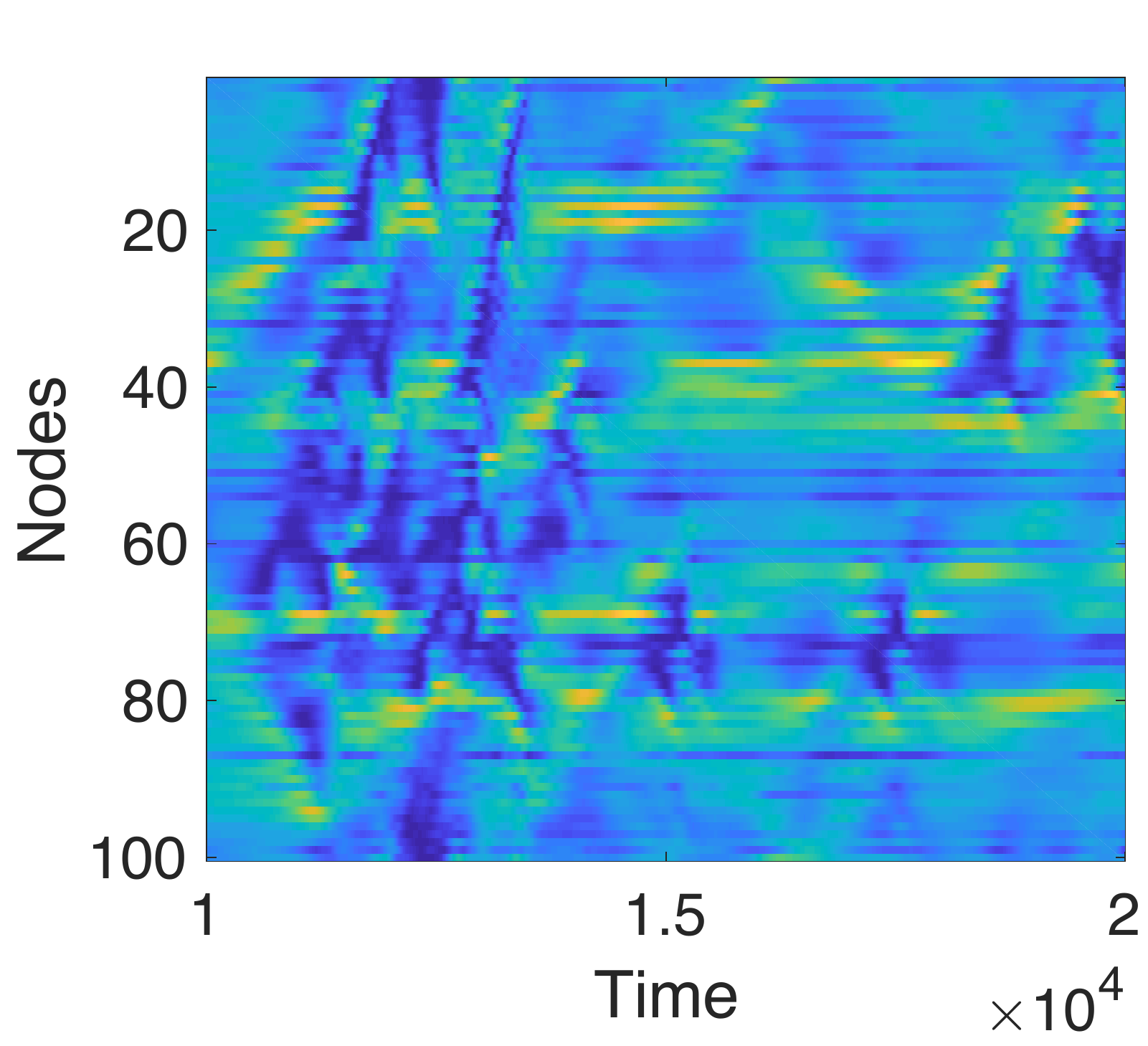}
	\label{1_sr_mod}}
	\hspace{2mm}
	\subfigure[]{\includegraphics[height=4.8cm]{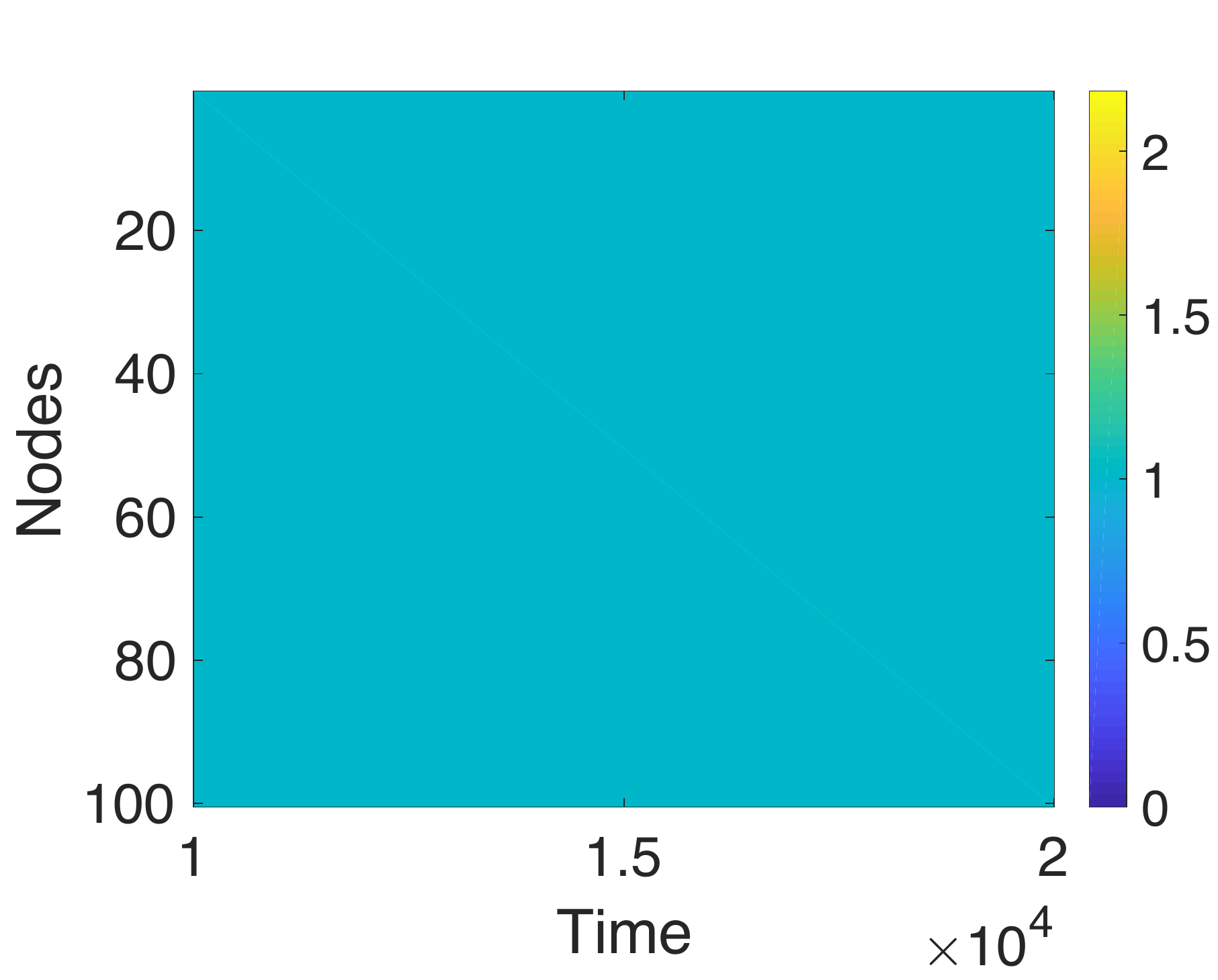}
	\label{1_lr_mod}}
	
	\caption{Instability and observed patterns for the model of  Stuart-Landau oscillators coupled via (i) a unidimensional lattice (first column) and (ii) two rewired versions of it, respectively  second  and third columns.  First row: dispersion relations and networks; second row: modulus of $|w_i|$ for every node, over a finite time window of observation. Self-loops are not represented in the network images. 
	}
	\label{fig_longrange1}
\end{figure}

The network displayed in the panel of fig.~\ref{fig_longrange1}(a) is a ring with $k=7$, including self-edges.  The associated dispersion relation is characterized by a collection of unstable modes that populate the region at small $|\Lambda^{(\alpha)}|$, where the instability localizes. The emerging patterns are plotted in Fig.  \ref{fig_longrange1}(d). For $\gamma=2$, the process of network generation is biased to short range couplings, see Fig. \ref{fig_longrange1}(b), and the  ensuing patterns are depicted in Fig. \ref{fig_longrange1}(e). When reducing $\gamma$, the spectral gap (the distance from zero of the first negative eigenvalue of the Laplacian) opens. For sufficiently small values of $\gamma$ ($\gamma=0.6<\gamma_c$, in Fig. \ref{fig_longrange1}(c)), all Laplacian eigenvalues are associated to negative entries of the dispersion relation. The fully synchronous state is hence stable and the imposed perturbation gets re-absorbed after a transient in time, see  figs.~\ref{fig_longrange1}(f).

\begin{figure}

	\subfigure[]{\includegraphics[width=4.5cm]{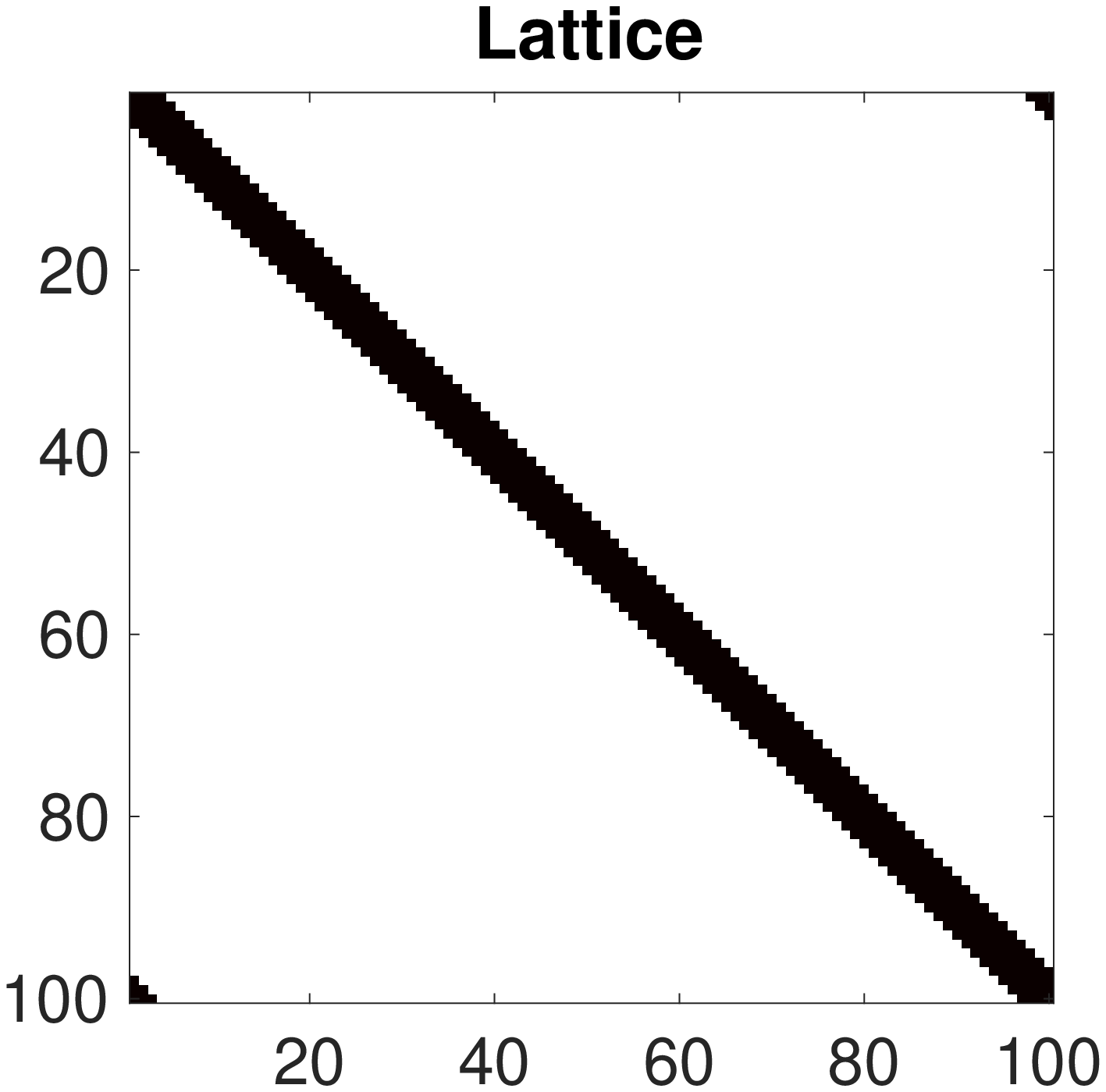}
	\label{1_lattice}}
	\hspace{10mm}
	\subfigure[]{\includegraphics[width=4.5cm]{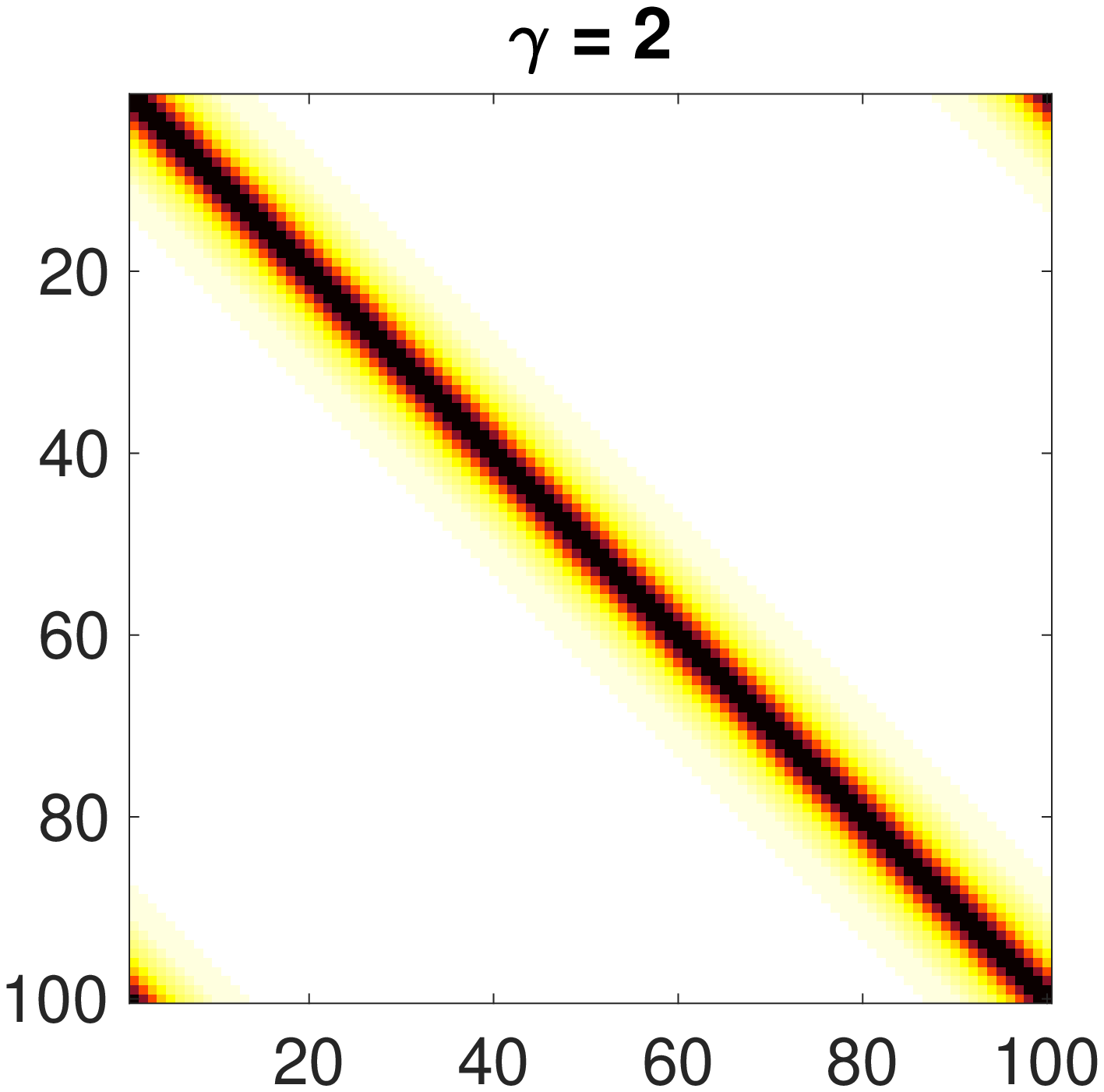}
	\label{1_shortrange}}
	\hspace{10mm}
	\subfigure[]{\includegraphics[width=5cm]{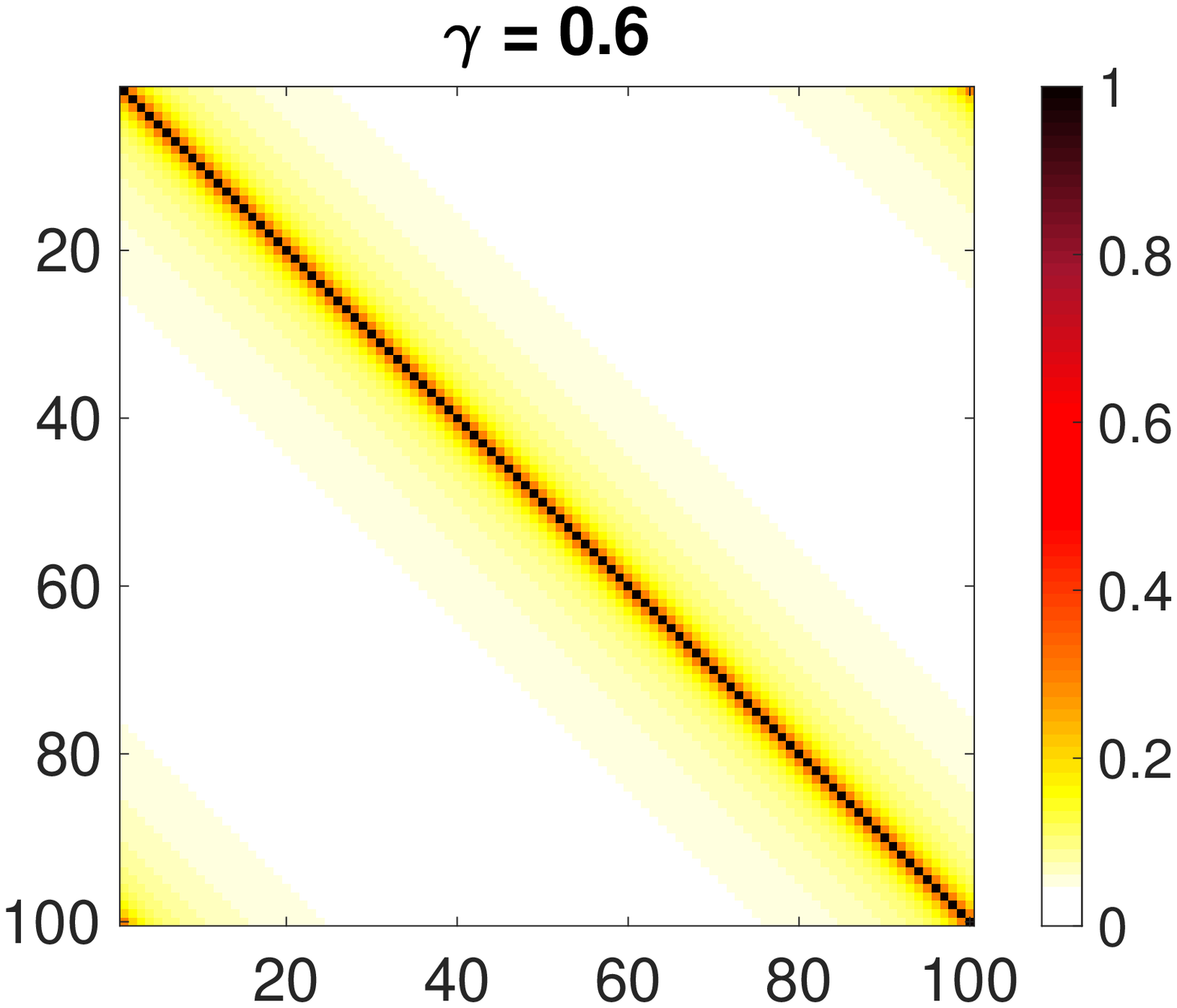}
	\label{1_longrange}}

	\subfigure[]{\includegraphics[width=6cm]{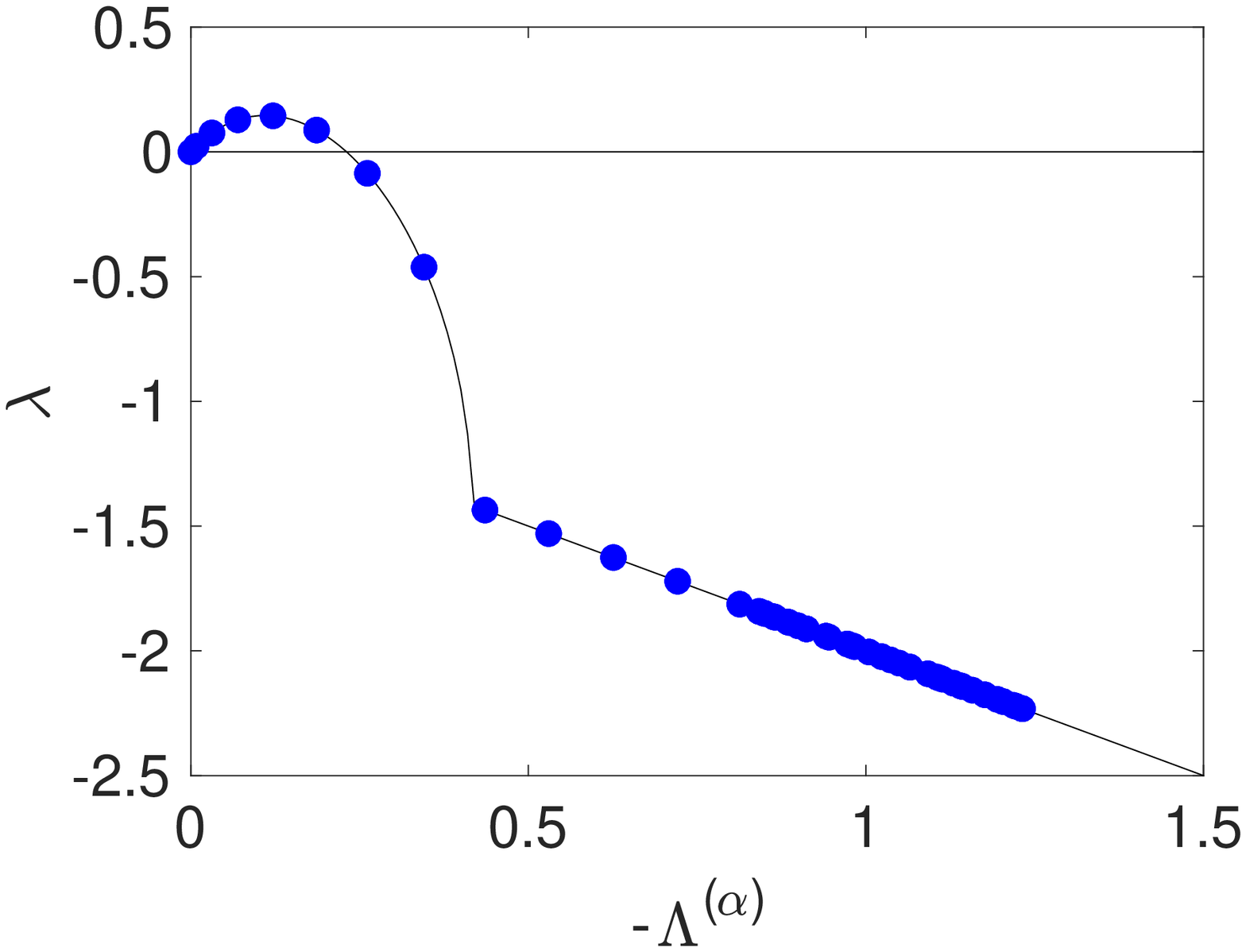}
	\label{1_lattice_dispRel}}
	\hspace{-5mm}
	\subfigure[]{\includegraphics[width=6cm]{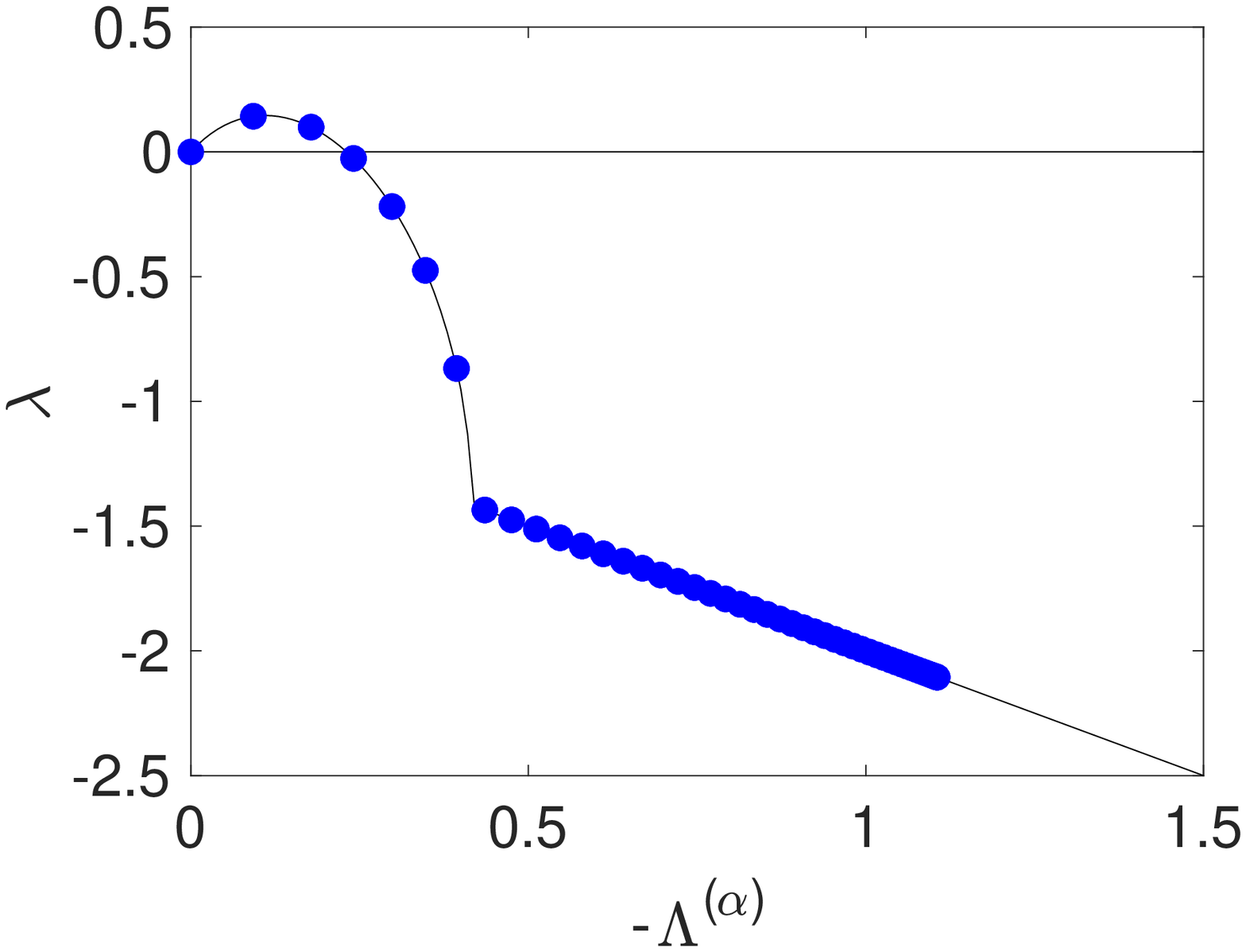}
	\label{1_sr_dispRel}}
	\hspace{-5mm}
	\subfigure[]{\includegraphics[width=6cm]{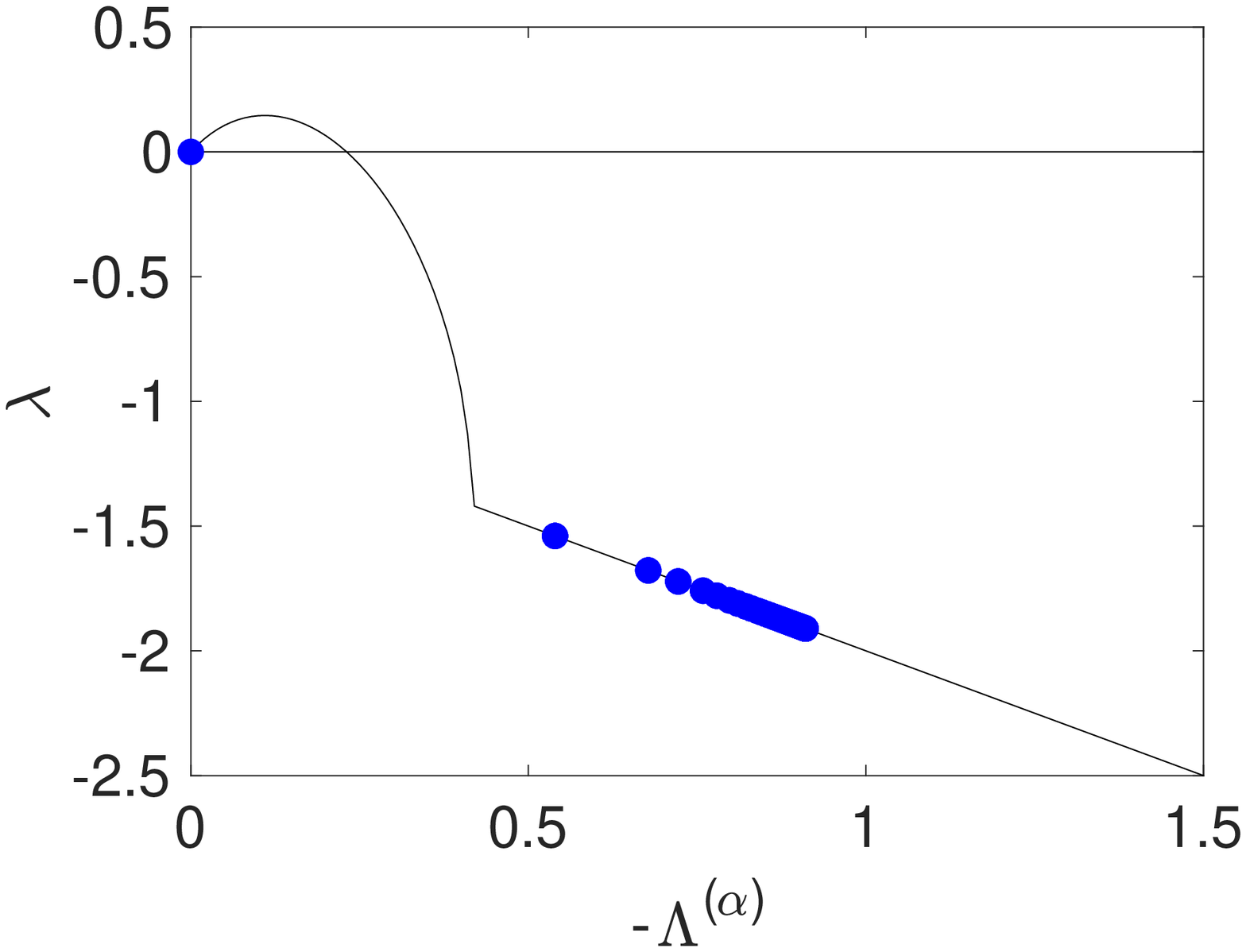}
	\label{1_lr_dispRel}}
	
	\subfigure[]{\includegraphics[height=4.5cm]{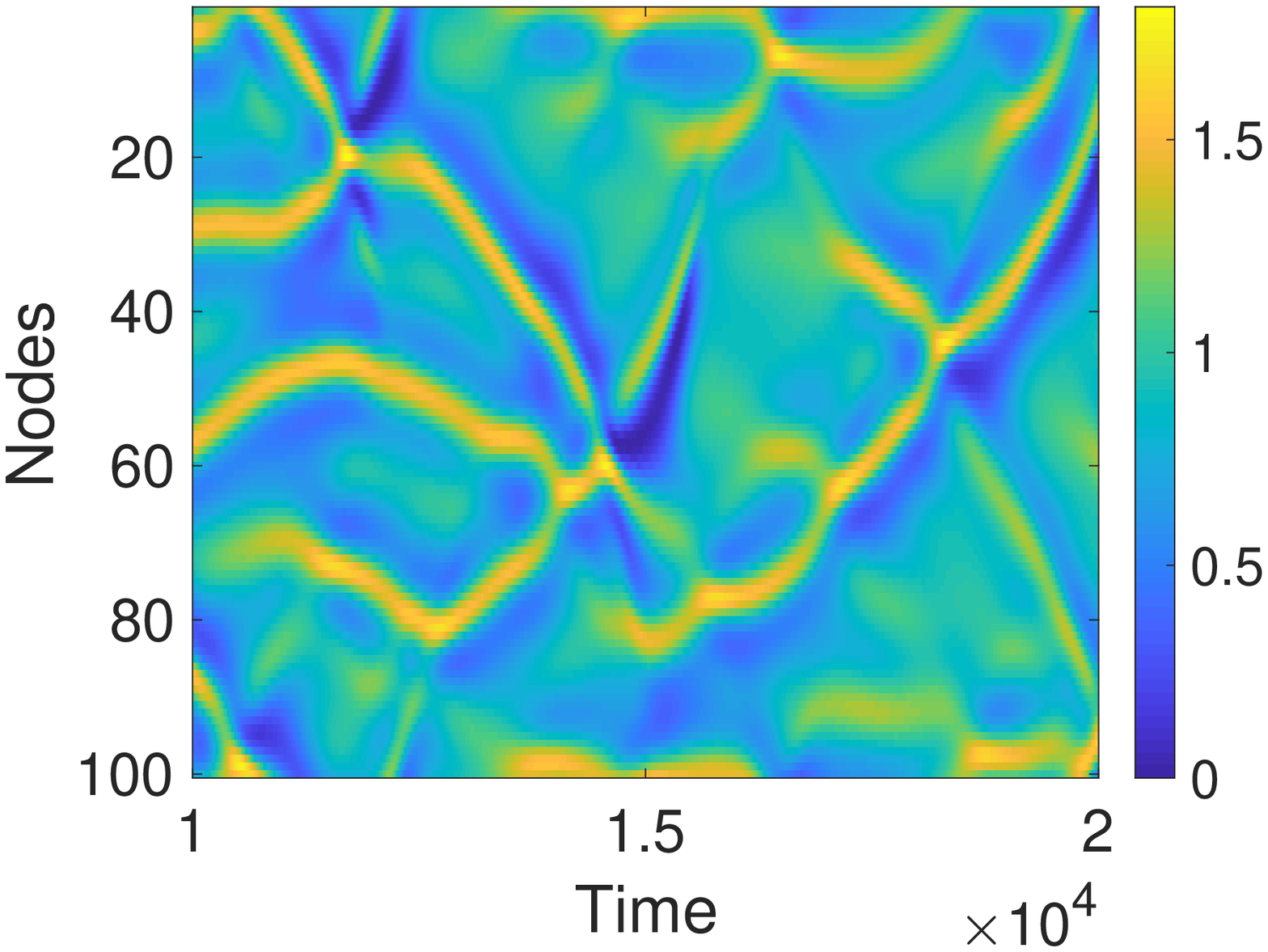}
	\label{1_lattice_mod}}
	\hspace{2mm}
	\subfigure[]{\includegraphics[height=4.5cm]{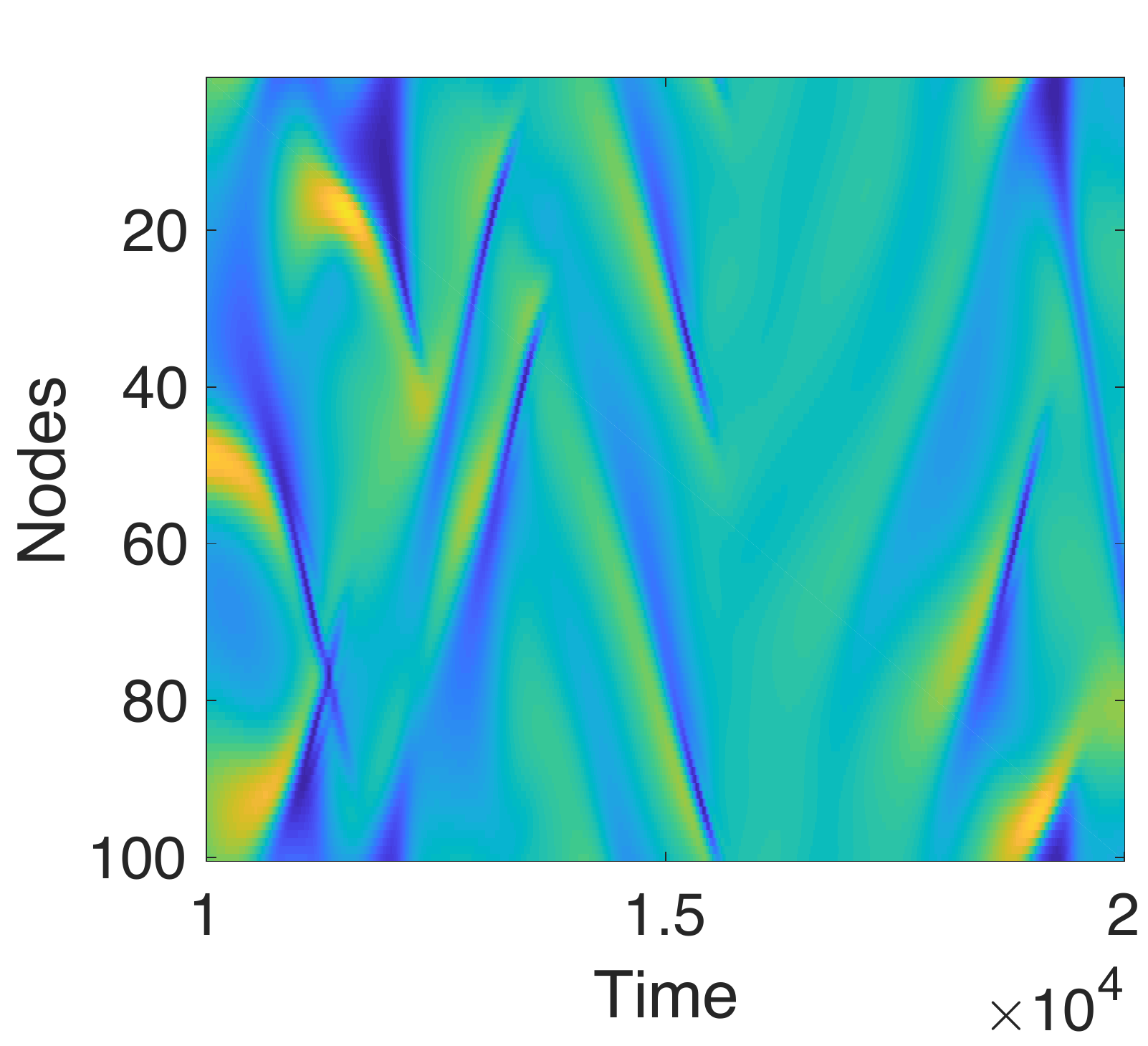}
	\label{1_sr_mod}}
	\hspace{2mm}
	\subfigure[]{\includegraphics[height=4.5cm]{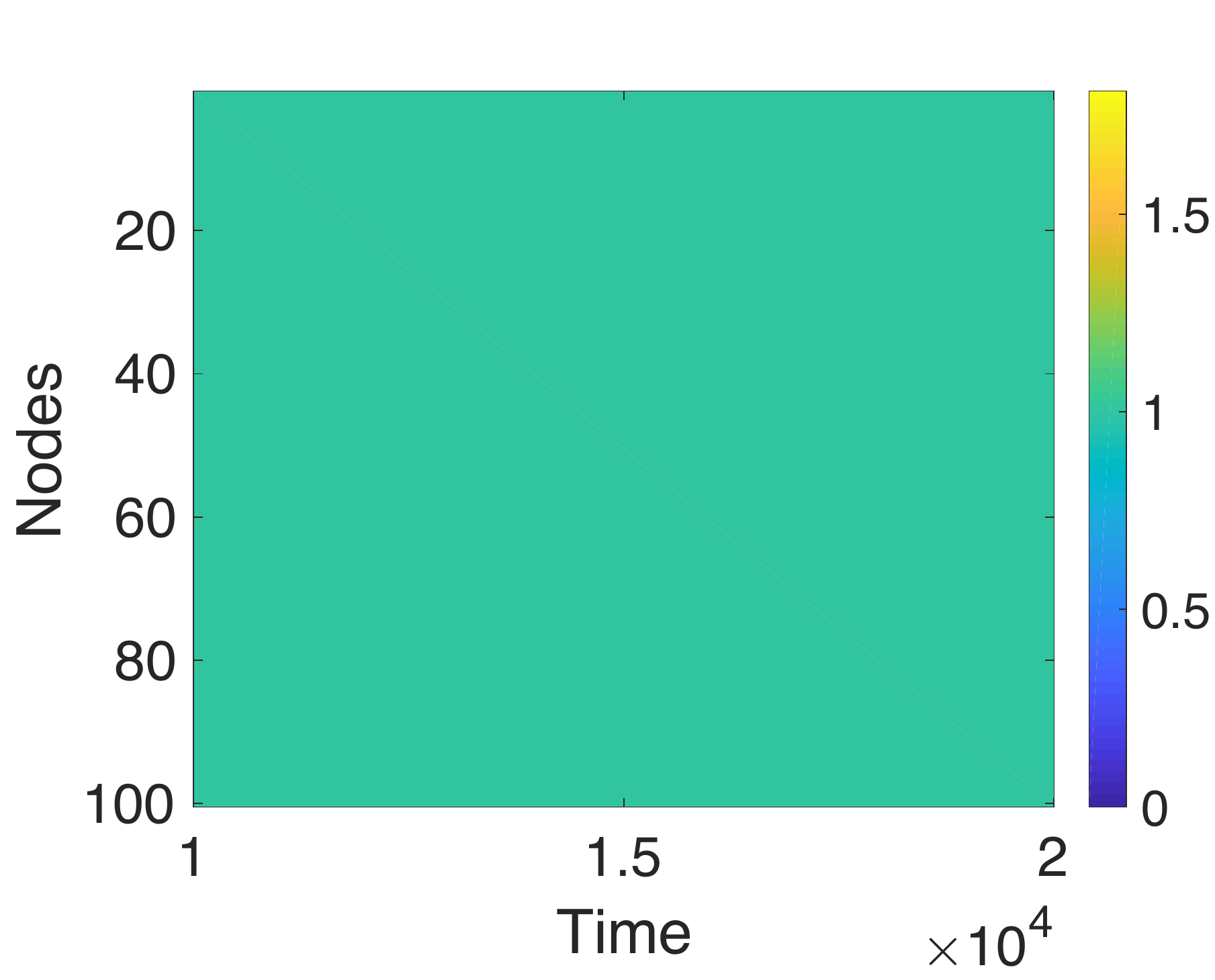}
	\label{1_lr_mod}}
	\caption{Instability and observed patterns for the model of  coupled Stuart-Landau oscillators. Here, the long-range effect is obtained by weighting the links in the adjacency matrix, according to a decaying power-law function with exponent $\gamma$. First row: scatter plot of the adjacency matrix weights, second row: dispersion relation, third row: modulus of $|w_i|$ for every node, over a finite time window of observation. 
	}
	\label{fig_longrange2}
\end{figure}

As a further check, we consider the Stuart-Landau oscillators coupled via a weighted adjacency matrix that implements the above strategy for interpolating from a short to a long-ranged scenario. The assigned weights scale as a power law of the nodes distance, with exponent $\gamma$. Stated differently, the dependence on the inter-nodes distance lies in the intensity of the connections. As clearly illustrated in Fig. \ref{fig_longrange2}, the spectral gap increases when progressively strengthening the connections among distant nodes (i.e. by decreasing $\gamma$), and the system gains a stable configuration characterized by the synchronous oscillators. By adding long-range couplings, one favors a mean-field homogenization of the dynamics, the system behaving hence as a tightly connected uniform component.


\end{document}